\documentclass{aa}  
\usepackage{graphicx}
\usepackage{natbib}
\bibpunct{(}{)}{;}{a}{}{,} 

\usepackage{txfonts}
\newcommand{\degree}{\ensuremath{^\circ}}
\newcommand{\Vhelio}{V$_{helio}$ }

\begin{document}

   \title{The interstellar  cloud  surrounding the Sun: a new perspective
   }

   \author{Cecile Gry
          \inst{1}
          \and
          Edward B. Jenkins \inst{2}
          }

   \institute{Aix Marseille Universit\'e, CNRS, LAM (Laboratoire d'Astrophysique de Marseille) UMR 7326, 13388, Marseille, France\\
              \email{cecile.gry@lam.fr}
         \and
             Department of Astrophysical Sciences, Princeton University Observatory, Princeton, NJ 08544, USA\\
             \email{ebj@astro.princeton.edu}
         }

   \date{Received 27 December 2013 ; Accepted 1 May 2014}
   
   \titlerunning{The interstellar cloud surrounding the Sun}
\authorrunning{C. Gry \& E.B. Jenkins}

 
  \abstract
   {}
   {We 
  offer a new, simpler picture of the local interstellar medium,  made of a single continuous cloud enveloping the Sun. This new outlook enables the description of a diffuse cloud from within and brings to light some  unexpected properties.
}
   {We re-examine the kinematics and abundances of the local interstellar gas, as revealed by the published results for the ultraviolet absorption lines of Mg~II, Fe~II, and H~I.}
   {
In contrast to previous representations, our new picture of the  local interstellar medium consists of a single, monolithic cloud that surrounds the Sun in all directions and accounts for  most of the matter present in the first 50 parsecs around the Sun. The cloud fills  the space around us out to about 9 pc  in most directions, although its boundary is very irregular with possibly a few  extensions up to 20 pc.
   The cloud does not behave like a rigid body: gas within the cloud is being differentially decelerated in the direction of motion, and the cloud is expanding in directions perpendicular to this flow, much like a squashed balloon.  
Average H~I volume densities inside the cloud vary between 0.03 and 0.1 cm$^{-3}$ over different directions. Metals appear to be significantly depleted onto grains, and there is a steady increase in depletion from the rear of the cloud to the apex of motion.  There is no evidence that changes in the ionizing radiation influence the apparent abundances. Secondary absorption components  are detected in 60\% of the sight lines. Almost all of them appear to be interior to the volume  occupied by the main cloud. 
Half of the sight lines exhibit 
a secondary component  moving at about $-7.2\,{\rm km~s}^{-1}$ with respect to the main component, which may be 
the signature of a shock propagating toward the cloud's interior.  
 }
   {}

\maketitle
   
%
%
\section{Introduction}

The study of the local interstellar medium (LISM) is interesting for two reasons: (1) an obvious aspect is the knowledge of the structure, the physical conditions, and the evolution of the gaseous environment surrounding our solar system, which allows us to understand better the interactions between the two and (2) the LISM is also an excellent laboratory for studying the basic physics at work in diffuse gas. Indeed, the simplicity of the short sight lines in the solar vicinity provides a unique opportunity to study individual regions, individual clouds, and individual interfaces that are usually blended in longer sight lines. 

The interstellar environment of the Sun, its  structure and physical conditions have been described with increasing precision in the last few decades, after space-based UV spectroscopy and soft X-ray observations have become available. For an overview of the present knowledge on the interstellar medium surrounding the Sun, see the comprehensive review by \cite{Frisch.Redfield.Slavin2011}.

The distribution of the local interstellar gas and its characteristics (velocity, abundances, temperature, turbulence, electron density, ionization fractions) are studied through the absorption lines it produces in the spectra of nearby stars.
From the low column densities found in most lines of sight up to about 100 pc, the Sun is known to be located in a region of particularly low interstellar gas density.  For several decades this region has been regarded as a volume filled mostly with a very low-density, very hot gas that emits a smooth background of soft X-rays, and thus it has been called  the Local Bubble. 
Whether or not this low density region is filled with hot gas is now subject to question, however the subject of this paper addresses the structure of the somewhat cooler gas that produces distinct velocity components observed in all short sight lines around the Sun.

All lines of sight observed so far in the UV, even the shortest ones, exhibit one or several velocity components of this warm,  low column density gas. Several studies have been conducted by different groups to characterize the local ISM gas responsible for these absorption components. Following the pioneering work of \cite{Crutcher1982}, \cite{Lallement1992}  have identified a coherent velocity vector creating a CaII absorption component in the spectra of six nearby stars located in the anti-Galactic center direction. Moreover, both investigators identified it with the gas of interstellar origin that 
flows into the heliosphere, producing the L$\alpha$ and He $\lambda$584 backscattering inside the solar system. \cite{Lallement1992} called the gas responsible for this absorption component the 'LIC' for local interstellar cloud. However, in the general direction of the Galactic center,  namely in the direction of $\alpha$ Cen (at a distance of 1.3 pc), they found that the velocity of the absorption feature failed to fit this common component  by a few ${\rm km~s}^{-1}$ difference in modulus. The interstellar gas  detected in the $\alpha$ Cen line of sight was therefore considered  to be a distinct cloud that was called 'G' for Galactic center cloud. 

Since then, there have been many more observations of the nearby interstellar gas using the CaII absorption lines in stellar spectra \citep{Frisch2002}, and these have been supplemented by observations in the UV, which features the ions Mg~II and Fe~II \citep{Redfield.Linsky2002} and the neutrals O~I and N~I \citep{Redfield.Linsky2004}  and H~I \citep{Wood2005}. All of these observations were used to study the nearby interstellar gas velocity distribution. 

\cite{Frisch2002} have  described the nearby interstellar medium within 30 pc of the Sun as a cluster of local interstellar cloudlets (that they call CLIC). They derived for the CLIC  a single "bulk" flow vector, with each cloudlet showing a velocity deviation of a few ${\rm km~s}^{-1}$ from the bulk flow. They showed that the CLIC is a decelerating flow, since they found a negative gradient in the velocity deviations from the bulk flow when going from upwind to downwind directions. With this description of the local ISM as a collection of cloudlets, \cite{Frisch.Mueller2011} proposed that the interstellar environment of the heliosphere, called  "Circum-Heliospheric Interstellar Medium" (CHISM),  might have changed in the last few 10\,000 years, or even in the last few 1000 years if the clouds are filamentary. 

\cite{Redfield.Linsky2008} (hereafter RL08) described an important attempt to organize the different components in terms of a  number of coherently moving 'clouds', each of which was identified using results that came from several lines of sight. They gathered the interstellar absorption information for 157 stars closer than 100 pc, of which 55\% have (partial) UV spectra. Using mostly kinematical and spatial information, these investigators created a dynamical model of the LISM that includes 15 different clouds that could be characterized by their locations on the sky and their velocity vectors. With these 15 distinct clouds, the model fits 81.2\% of the velocity components. In the interpretation of RL08, the LIC covers about 44\% of the sky and the G cloud about 13\%.  From the derived velocities of the clouds they stated that the heliosphere is not immersed in the LIC, but instead is situated in a transition region between the two clouds. 

The model of RL08  raises a few questions. In particular, with such a configuration in which  the Sun is surrounded by several distinct and separate clouds, we would expect to find some lines of sight going through the gaps between the clouds, which would create cases showing no absorption. However such a line of sight has never been found: every line of sight observed with a sensitive enough absorption line in the UV shows at least one absorption component. Moreover,  \cite{Crawford1998} failed to find evidence to support the existence of the G Cloud as a separate entity, despite very high-resolution observations of interstellar Ca II K lines toward sight lines where both clouds were expected.

We therefore decided to re-examine the LISM UV-absorption database with a new focus, i.e., making the assumption that the CHISM can be assimilated to a unique, continuous medium that envelops  the Sun in all directions and produces one
absorption component in every line of sight. With this assumption in mind, we examine to what extent that medium can be assimilated to a unique interstellar cloud, 
and what limitations this implies to the usual picture of a homogeneous cloud, moving as a rigid body.     

\section{Identification of a single circum-heliospheric cloud from Mg~II and Fe~II kinematics}
\subsection{The database}
In this paper we examine   the UV Mg~II and Fe~II data. 
The reasons for choosing these lines are the following:\newline
- Mg~II and Fe~II produce UV absorption lines that are among the most sensitive to detect small column densities of matter. \\
- These elements are observed in a large number of sight lines because their lines are located in the near UV, where cool stars still produce a continuum with a good signal-to-noise ratio or have stellar emission lines that can serve as continua for the ISM lines (as opposed to the situations for H~I, O~I, N~I, and C~II, which have lines that are below 1500 A).\newline 
- Since they all come from observations taken by the Hubble space telescope (HST), the data and their processing are easily traceable, and in particular their wavelength calibrations are more likely to be consistent, which is an important aspect for a study based on kinematic properties.\newline
- They are among the heaviest elements (especially Fe~II) observed in absorption. This makes their absorption lines the narrowest and therefore less subject to blending when multiple components with small separations are present. \newline
- We use only high resolution data (R$\simeq$100\,000) to enable the best differentiation of velocity components and provide the most accurate velocity values. 
\par We have chosen not to include the Ca~II measurements in the kinematical study for two reasons. The first is the sensitivity difference between Ca~II lines and the UV lines. In the few lines of sight  observed in both Ca~II and the UV, a number of components detected in the UV are not seen in Ca~II. Therefore, to guarantee the homogeneity in the sensitivity of the sample, we prefer to limit the data to the most easily detected lines of Fe~II and Mg~II in the UV. Second, we have noticed  differences of up to 3 ${\rm km~s}^{-1}$ in velocities between different (N, V, b) solutions found by different authors for Ca~II or Na~I for a same sight line (reported by \cite{Redfield.Linsky2002}).
These differences might be due to the use of different instruments, different wavelength calibrations, or different spectral analysis methods. In any case, since our intention  is not to redo any measurement but to use existing material, in these cases we do not know which value to choose. Therefore, to preserve the integrity of the velocity calibration of the sample, we decided to leave  out the CaII and NaI samples. 
\par Our database is  drawn from \cite{Redfield.Linsky2002}'s Table 3  (for Mg~II) and Table 4 (for Fe~II), which give for each target the velocity, broadening parameter, and column density of each component. \\
Since a new wavelength calibration of the Goddard High Resolution Spectrograph (GHRS) \citep{Kamp2006} has been released after the publication of \cite{Redfield.Linsky2002}'s effort, we have checked all GHRS data sets used in their Tables 3 and 4 to see if the most recent versions of the data in the archive were consistent with the published spectra. When new data from the Space Telescope Imaging Spectrograph (STIS) were available for a target, we compared them with the published spectra. In a large majority of cases, the concordance was very good.  Thus, we only had to make minor corrections to a total of six sight lines, all of which had a magnitude less than the formal wavelength precision of the GHRS. We did not reanalyze the data, but instead we chose  to adopt the measurements published by \cite{Redfield.Linsky2002}, with the consequence that except for the few small velocity adjustments mentioned above, our input data base is practically identical to that used by RL08. This should facilitate any comparisons of the findings. 

Our sample includes a total of 111 components, in a total of 59 lines of sight, of which 54 are observed in Mg~II, 32 are observed in Fe~II, and  27 are observed in both elements.

\subsection{Identifying the CHISM  candidate component in each line of sight \label{sec:select}}
Our first step was to identify in every line of sight of the database the  component that is most likely to originate in the  Circum-Heliospheric ISM (CHISM). 
For this purpose,  we calculated for each line of sight $i$ the quantity V$_{i,helio}$, the projection onto the line of sight of the velocity vector  $\overrightarrow{V_0} $ of the ISM flow interacting with the heliosphere. If the Sun were completely surrounded by a contiguous interstellar medium moving as a rigid body and not rotating, then in every line of sight we would expect to see  an absorption component at the appropriate value of V$_{i,helio}$.\\
 The velocity vector $\overrightarrow{V_0} $ of the ISM flow into the heliosphere has been measured inside the solar system by various experiments. In particular, \cite{Moebius2004} offered a synthesis of the measurements made by Ulysses \citep{Witte2004} and previous experiments, and \cite{McComas2012} provided a new, slightly different   velocity vector derived from the satellite IBEX:\\
$\overrightarrow{V_0} $ by Ulysses is defined by \[V_0 = 26.24 \pm0.45~{\rm km~s}^{-1}, \,l_0 = 183.36\pm0.54\degree, \,b_0=-15.94\pm0.62\degree,\]
while $\overrightarrow{V_0} $ from IBEX is defined by  \[V_0 = 23.2 \pm0.3~{\rm km~s}^{-1}, \,\,l_0 = 185.25\pm0.24\degree, \,b_0 =-12.03\pm0.51\degree.\]
For each  line of sight defined by the target Galactic coordinates ($l_i$,$b_i$), we derive V$_{i,helio}$ by the following: \[
V_{i,helio}=V_0 \,(\cos b_i  \,\cos b_0 \,\cos (l_0-l_i) + \sin b_0 \,\sin b_i) \]with V$_0$, $l_0$ and b$_0$ being
 the modulus, Galactic longitude, and Galactic latitude of either of the two vectors $\overrightarrow{V_0} $ given above. 
 We can then  compare V$_{i,helio}$ with the velocity of all components in each line of sight and designate <<Component~1>> the CHISM candidate, i.e., the component whose velocity is closest to V$_{i,helio}$.  
 We have found that the two V$_{helio}$'s derived with either of the two velocity vectors are close enough to each other that they both define the same Component~1 in all lines of sight.
 By definition, all lines of sight observed in either element give rise to a "Component~1"\footnote{except for $\alpha$~Tau and $\gamma$~Cru mentioned by \cite{Redfield.Linsky2002}, because  their spectra show no flux at the location of Component~1 due to deep P-Cygni profiles, thus preventing any measurement.}, irrespective of its velocity agreement with V$_{i,helio}$.
Our ``Component~1"  sample is thus made of 59 components in total, including 54 components observed in Mg~II, 32 components observed in Fe~II, and  27 components observed in both lines. 
\subsection{Component~1 velocities relative to \Vhelio \label{sec:residuals}}
\begin{figure}[!htbp]
\includegraphics[width=1.0\columnwidth]{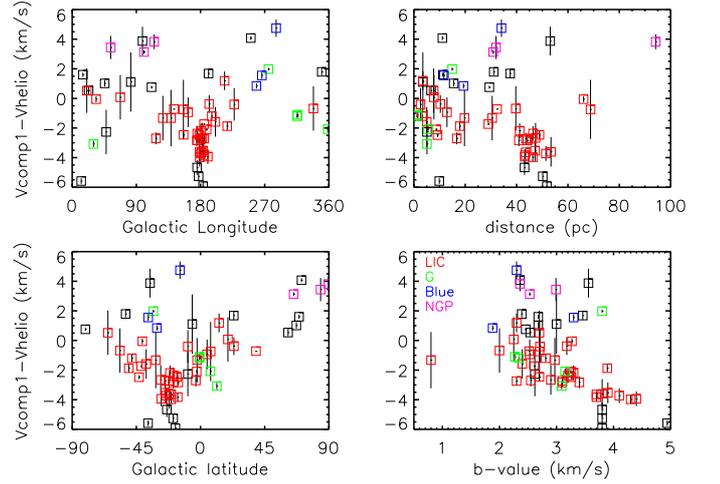} 
\caption{Measured velocity deviations relative to \Vhelio for Component~1, the component that is closest to \Vhelio in each sight line.  The error bars correspond to the error on the determination of V$_{i,1}$ \citep{Redfield.Linsky2002}. The colors indicate the  components  assigned to the LIC (red), G cloud (green), Blue Cloud (blue) and NGP cloud (pink) in the RL08 model, i.e., the four RL08 clouds that include at least three of our Components~1. In this figure, V$_{i,helio}$ has been calculated from the Ulysses vector.  The electronic version is in color.
\label{fig:residuals} }
\end{figure}

The consistency of the sample of Components~1 with the representation of a single cloud surrounding the solar system can be appraised in a first step by 
looking at the velocity discrepancies  
 V$_{i,1}$- V$_{i,helio}$ where V$_{i,1}$ is the observed Component~1 velocity for line of sight~$i$. The velocity discrepancies  calculated with the Ulysses vector are shown in Figure~\ref{fig:residuals}, plotted versus several quantities. In this figure we use the MgII velocity when it is available and the FeII velocity otherwise. \\ 
The error bars shown in this figure correspond to formal errors listed in \citep{Redfield.Linsky2002}, which for some of them  reflect only the accuracy of velocity measurements from a calibrated spectrum but underestimate the uncertainty of the GHRS wavelength calibration ($\simeq$~1~${\rm km~s}^{-1}$ when both proper observing and data reduction techniques are used).  One can get a good idea of 
the global velocity measurement uncertainties by considering the 
differences between the measured values of V$_1$(Mg~II) and V$_1$(Fe~II). Those can differ because the Mg~II and Fe~II lines have been analyzed separately for most lines of sight. Since Mg~II and Fe~II are very likely to originate in exactly the same medium,  their differences hence indicate the velocity determination errors due to both wavelength calibration discrepancies and spectral analysis uncertainties. 
\cite{Redfield.Linsky2002} quote a standard deviation of 1.1~${\rm km~s}^{-1}$ for the Mg~II and Fe~II velocity differences (their distribution is shown on their Figure~4). 
Therefore, we conclude that measurement uncertainties  are of the order of 1.1~${\rm km~s}^{-1}$ (1-$\sigma$). 
On the full Components~1 sample the velocity discrepancies have a mean value of -1.0~${\rm km~s}^{-1}$ and a dispersion of 2.5~${\rm km~s}^{-1}$, by consequence  velocity uncertainties  account for almost half  the dispersion of the  discrepancies in the whole sample.\\
From the examination of Fig.~\ref{fig:residuals} we can make the following observations:

1) Deviations larger than expected from the measurement uncertainties 
occur in a couple of special cases and in two particular regions:\\
-   One of the largest negative deviations correspond to AU~Mic. Its $b$-value of 5~${\rm km~s}^{-1}$ is the highest of the whole sample and its spectral profile \citep{Redfield.Linsky2002} shows evidence for the presence of two components whereas it was fitted with a single component, implying that the accuracy of the velocity measurement is compromised. Moreover, since one of these components is possibly associated with the debris disk that is known to exist around AU~Mic \citep{Kalas2004}, this star will be removed from the sample for the rest of the analysis.\\
-   One of the  largest positive deviations  corresponds to  $\upsilon$~Peg. In this case  the velocity determination is also very uncertain because the Component~1 absorption line occurs in the wing of a much stronger component. The uncertainty is indeed revealed by Figure~1 in \cite{Redfield.Linsky2002}, where the plotted fits for Mg~II and for Fe~II exhibit a difference of 4~${\rm km~s}^{-1}$. As a  consequence, $\upsilon$~Peg  will also be removed from the Component~1 sample for the kinematic study.\\
-	The other largest negative deviations, which extend below $-4~{\rm km~s}^{-1}$, are  found in the lines of sight toward members of  the Hyades Cluster around $l=180$\degree\ and $b=-20$\degree, which is very close to the apex of $\overrightarrow{V_0} $. In this direction all components that have velocity discrepancies below $-$3~${\rm km~s}^{-1}$ have $b$-values above 3.6~${\rm km~s}^{-1}$,  showing that extra motions are associated with those components. In fact, for components showing negative velocity deviations, there is a marked trend of increasing $b$-values with increasing deviations. 
One may wonder if the  high velocity discrepancies and the high $b$-values of the interstellar components in the direction of the Hyades Cluster  could be related to perturbation due to the wind of the cluster stars, although these stars are located 40~pc away, hence some 30~pc away from the local interstellar cloud(s).  This trend could alternatively be the signature of a velocity gradient within the cloud already noticed by \cite{Redfield.Linsky2001}.\\
-	Four of the six  highest positive deviations --higher than +3 ${\rm km~s}^{-1}$-- correspond to lines of sight toward  high positive Galactic latitude objects. In the northern Galactic hemisphere, the deviations increase from $\sim-2$ to $\sim+4$ ${\rm km~s}^{-1}$ for Galactic latitude $b$ increasing from 0\degree\ to 90\degree. This may suggest a slight error in the determination of the latitude projection of the heliospheric velocity vector or, here again, the presence of a velocity gradient when moving up toward the north Galactic pole.

2) We also note that outside the two above mentioned regions, 
there is no  substantial difference between the velocity discrepancies of the components that have been assigned to the LIC in the RL08 model (average deviation of -1.9~${\rm km~s}^{-1}$ with a dispersion of $1.4\,{\rm km~s}^{-1}$) and the discrepancies of the components that have been assigned to another cloud. This is true in particular for the G~Cloud (in green in Figure~\ref{fig:residuals})  the components of which have an average deviation of -1.1~${\rm km~s}^{-1}$ with a dispersion of $1.9\,{\rm km~s}^{-1}$. We note that the $\alpha$~Cen component, 
which triggered the introduction of the G~Cloud in the first place, has a modest discrepancy of  $-1.1~{\rm km~s}^{-1}$,  equal to 1-$\sigma$ of the  measurement errors (as represented by the difference of MgII and FeII velocity measurements) and comparable to the discrepancy of most components assigned to the LIC. 
Having said that, when we plot the velocity deviations relative to \Vhelio calculated with the IBEX vector instead of the Ulysses vector, the largest negative velocity discrepancies occur for targets around $l$=0\degree\ and $b=0$\degree, i.e.,  close to the anti-apex of $\overrightarrow{V_0} $, corresponding to three of the  G cloud representatives.

In conclusion, most lines of sight in all directions include a component that is within the velocity measurement errors 
from the velocity vector $\overrightarrow{V_0} $ of the ISM flow into the heliosphere.
Exceptions to this include only sight lines toward  the apex or anti-apex of $\overrightarrow{V_0} $, where  negative velocity deviations are present, and  sight lines at high positive Galactic latitudes (i.e., perpendicular to $\overrightarrow{V_0}$), where  positive velocity deviations are observed. 
\subsection{Mean velocity vector fitting the Component~1 database.\label{sec:velvectors}}
We then derived the velocity vector  that best fits the Component~1 sample, i.e., the velocity vector $\overrightarrow{V_m} $ --expressed in the heliocentric reference frame and defined by $l_m$, $b_m$, V$_m$-- which minimizes the quadratic sum of the differences between  the projections of $\overrightarrow{V_m} $ onto the lines of sight ($l_i$, $b_i$)) and the measured velocities V$_{i,1}$:
$\Sigma_i(V_m[\cos b_i\,\cos b_m\,\cos(l_m-l_i) + \sin b_m\, \sin b_i] - V_{i,1})^2$. Since all but five Fe~II sight lines are included in the Mg~II sample, we kept the two samples separate and performed the fit on the most numerous Mg~II data.
When considering all 52 Mg~II Components~1 (AU~Mic and $\upsilon$~Peg were eliminated from the sample for the reasons stated in Section~\ref{sec:residuals}), the best fit was found as follows: \[l_m=185.33\pm0.86\degree ; \,b_m=-12.42\pm0.70\degree ;V_m=24.22\pm0.19{\rm km/s}.\]
We note however that 17 sight lines out of 52 in the Mg~II sample are in the direction   of members of the Hyades Cluster. This results in one third of the sample having almost identical $l$, $b$, and V$_1$. As a result, there is a very strong weight to this particular direction, and this may introduce a significant  bias in the sample. 
We therefore performed another fit after replacing  the 17 Hyades sight lines  by one virtual sight line whose characteristics are an average of the characteristics of the 17 Hyades sight lines. The fit now gives:\[l_m=185.84\pm0.83\degree ; \,b_m=-12.79\pm0.67\degree ;  V_m=25.53\pm0.26 {\rm km/s}.\]
The direction of the velocity vector has hardly changed, but the velocity scalar is 1.3 ${\rm km~s}^{-1}$ higher than when all Hyades sight lines are included, which is logical since the Hyades Cluster is in the apex direction, where the velocity is lowest.
{\footnotesize
   \begin{table}[!htbp]
      \caption[]{Comparison of the velocity vectors
         \label{tab:velvectors}}
         \begin{tabular}{lccc}
            \hline 
   &   $l$ (\degree)&    $b$ (\degree)&   V (${\rm km~s}^{-1}$)\\
               \hline 
 \multicolumn{4}{l}{\bf Heliocentric reference frame}  \\            
Ulysses vector &183.36$\pm$0.54&$-15.94\pm0.62$&26.24$\pm$0.45\\
 IBEX vector &  185.25$\pm$0.24&$-12.03\pm$0.51&23.2$\pm$0.3\\
comp1 mean vector$^{a}$ &185.33$\pm$0.86&$-12.42\pm$0.70&24.22$\pm$0.19\\
comp1 mean vector$^{b}$ &&&\\ 
 (origin for $\theta_m$)
&185.84$\pm$0.83&$-12.79\pm$0.67&25.53$\pm$0.26\\
 \hline 
  \multicolumn{4}{l}{\bf LSR reference frame}\\
  Mean comp 1$^{b}$ &169.53$\pm$1.38&5.81$\pm$1.10&15.08$\pm$0.27\\
  \hline
  \multicolumn{4}{l}{{\bf Cloud reference frame }(Section~\ref{sec:deform})}\\
  Cloud minor axis: &&&\\
 ( origin for $\theta_d$) 
 &174.2 $\pm$5.2  &$-12.1\pm$4.2 &\\
  \hline
  \hline
\end{tabular}
$^{a}$ {\footnotesize \it Mean velocity vector derived from all Components~1}\\
$^{b}${\footnotesize \it Mean velocity vector derived from Components~1, after replacing the 17 Hyades sight lines  by one virtual average sightline}
\end{table} 
}
Our  mean velocity vectors are given in Table~\ref{tab:velvectors}, where they are compared to the vectors of the interstellar gas entering the heliosphere measured either by Ulysses and previous experiments \citep{Moebius2004} or by IBEX \citep{McComas2012} . These two heliospheric vectors are slightly different from one another. \cite{Frisch.Science2013} have interpreted this difference in terms of a real, small-scale time variation induced by the motion of the  solar system through a turbulent  ISM  whereas the reality of the difference has been questioned by \cite{Lallement.Bertaux2014}. 
At any rate, the potential time variations are  
 not relevant for the absorption line observations used here since the small-scale inhomogeneities due to turbulence are averaged out on the long sight lines. Our measured velocity vector is indeed very close to both measurements: it has the same direction as the IBEX vector, but the velocity amplitude is closer to that of the Ulysses measurement.

We have also estimated the mean velocity vector in the local standard of rest (LSR), in order to express it independently of the Sun's velocity. We have adopted the LSR solar motion derived with Hipparcos data by \cite{Dehnen.Binney1998}, which has been translated  
into a velocity of 13.4 ${\rm km~s}^{-1}$ toward l=27.7\degree , b=32.4\degree\ by \cite{Frisch.Slavin2006}. 
The resulting mean LSR velocity vector is given in Table~\ref{tab:velvectors}:
its direction is about 25\degree\ away from the heliocentric direction of motion, and its modulus is about 10~${\rm km~s}^{-1}$ lower than in the heliocentric system.

We define the angle $\theta_{i,m}$, the angular distance of a line of sight ($l_i$, $b_i$) from the apex direction ($l_m$,$b_m$) of the mean velocity vector  $\overrightarrow{V_m} $:\[ \theta_{i,m} =  \cos^{-1}
(\sin\,b_i\,sin\,b_m+\cos\,b_i\,\cos\,b_m\,\cos(l_i-l_m)).\] 
If our Component~1 sample represents a homogeneous cloud surrounding the Sun in all directions and moves coherently according to the above velocity vector, the velocities of all Components~1 will be equal to V$_m \cos\, \theta_{i,m}$, and will be aligned on a perfect cosine curve.
\begin{figure}[!htbp]
\begin{tabular}{cc}
\includegraphics[width=0.95\columnwidth]{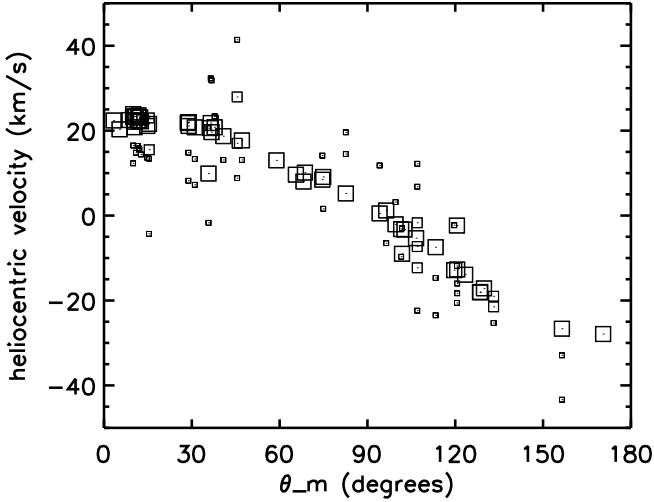} 
\end{tabular}
\caption{Measured velocities  of all components plotted against their corresponding angles $\theta_{i,m}$ away from the apex of the best-fitting velocity vector for Components~1. Large symbols indicate components that are dominant in their line of sight in terms of column density. Medium-size symbols depict cases where two secondary components in a line of sight have identical column densities within their error bars. The smallest symbols indicate components that are clearly secondary.  The adopted reference column densities are those of Fe~II, when available, and Mg~II otherwise. Note that nearly all the largest symbols are aligned with a curve that we can identify with the overall trend V$_m \cos\, \theta_{i,m}$ of Components~1,  where V$_m=25.53\,{\rm km~s}^{-1}$. \label{fig:velocity-domine} }
\end{figure}
 Figure~\ref{fig:velocity-domine} shows the velocities  of all components against 
$\theta_{i,m}$. In each case, the symbol size indicates whether or not a component is dominant  in the line of sight in terms of column density. It is clear that there is a conspicuous cosine curve, revealing that Components~1 are distinguishable from the other components in terms of an expected behavior in velocity. Moreover, this plot also demonstrates that for almost every line of sight, Component~1 is the dominant component. 
Thus, while our choices for Component~1 are based on a preconceived idea about the velocity of a single cloud, they are supported by a dominance over other components in sight lines that exhibit multiple components. 

Note that Figure~\ref{fig:velocity-domine} is much like the merging of Figures~3-a and 3-b of \cite{Frisch.Redfield.Slavin2011} that showed the bulk motion of the LISM found by \cite{Frisch2002} and the components identified with the LIC in RL08.  However, our plot differs from these because we calculated the mean vector after selecting one and only one component (Component~1) per line of sight, thereby tentatively defining a unique cloud that surrounds the Sun. 
Our selected Components~1  include most of the components previously identified with the LIC, but they also include additional  components that had been previously assigned to distinct clouds. The latter represent 40\% of the Components~1 sample.

\begin{figure}[!htbp]
\begin{center}
\includegraphics[width=0.95\columnwidth]{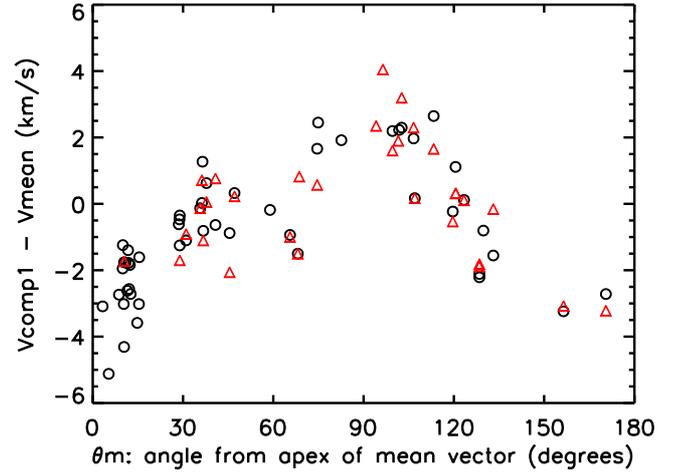}
\end{center}
\caption{Velocity residuals of Components~1, the components closest to \Vhelio, relative to the best-fit vector derived from the heliocentric Mg~II velocities  after replacing all Hyades sight lines by their average. Black circles: Mg~II, red triangles: Fe~II
\label{fig:VMgII} }
\end{figure}

The residuals of the Mg~II and Fe~II radial velocities relative to our  best-fit mean vector $\overrightarrow{V_m} $ are shown in Figure~\ref{fig:VMgII}   as  functions of $\theta_m$, the angular distance away from the pointing direction of this vector. 
This plot shows a clear trend with $\theta_m$:
The residuals increase from negative values at $\theta_m=0$ to positive values up to  $\theta_m$ = 90\degree\ and decrease again up to $\theta_m$ = 180\degree. The velocities are lower than the mean velocity toward the apex and anti-apex directions, and higher than the mean velocity in the directions perpendicular to the motion.\\
To check the reality of this dependence, we have explored the space of parameter variations, i.e., we have slightly varied  each of the parameters of the best-fit velocity vector represented by l$_m$, b$_m$ and V$_m$ and analyzed the resulting residuals.
 Although the general shape varies slightly with each parameter variation, the dependence of the residuals against $\theta_m$ (increasing and later decreasing) is always present. This is a strong indication that this trend is real and not simply an artifact of having an incorrect choice for $\overrightarrow{V_m}$. 

\subsection{A unique cloud surrounding the Sun \label{sec:unique}}
The trend shown in Fig.~\ref{fig:VMgII} has an interesting consequence. The simple, symmetrical single hump pattern seen here indicates this is a fundamental, second-order dynamical effect associated with an otherwise coherently moving cloud.
In effect, when this trend is taken into account as a systematic velocity gradient relative to  $\theta_m$, the remaining dispersion of the velocity residuals 
for all lines of sight is now reduced to the level of  the  uncertainties in the velocity determinations that we identified in Section~\ref{sec:residuals}. This means that the velocity information of our Components~1 sample can indeed be interpreted as the projected velocities of a unique interstellar cloud, the velocity of which corresponds to the ISM flow into the heliosphere, 
provided we allow for small, systematic changes of velocity that are internal to this cloud. Since all sightlines give rise to a Component~1, this shows that  the Mg~II and Fe~II data support the existence of a coherently moving, although not completely rigid,  interstellar cloud that surrounds the Sun in all directions. 

Our findings have implications about the morphology of the local ISM. We advocate the viewpoint that instead of being surrounded by several distinct clouds in our immediate vicinity, the solar system may instead be completely embedded in a continuous cloud that exhibits some relatively uniform internal motions in addition to its overall movement in space. 
We note that most (34 out of the 36) components assigned to the LIC in the RL08 model are identified with our Components~1 as expected, but  six out of the seven components assigned  by RL08 to the G cloud, as well as all components assigned to the clouds NGP, Leo, Aur, and Cet, are also tagged as Components~1 (see the Appendix for details).  This means that the domains identified with both the LIC and the G cloud, as well as a number of other clouds that had to be invoked to account for slight velocity deviations,  are now seen to be included in this single circum-heliospheric cloud.

Now that we have constructed a more simple outlook on the nature of the gaseous material that surrounds the solar system, we can move on to examine what the existing database can tell us about the structure and other interesting properties of this Local Cloud.

\section{Dynamics within the cloud} 
\subsection{Spatial distribution of the velocity distortions}
\begin{figure}[!htbp]
\begin{center}
\includegraphics[width=0.9\columnwidth]{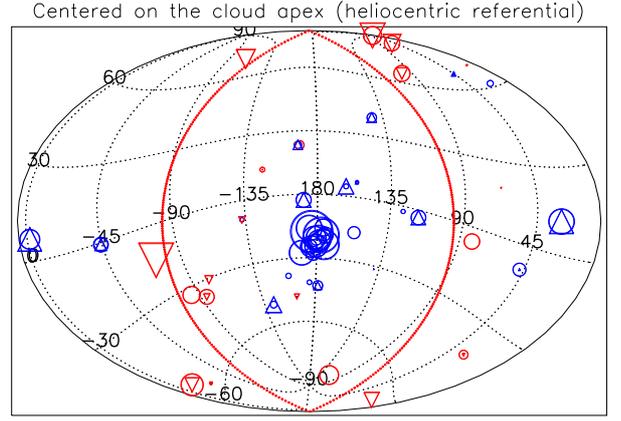}
\end{center}
\caption{Velocity deviations from the best-fit velocity vector, plotted against Galactic coordinates and centered on the apex of this vector. The red curve represents directions where $\theta_m=90\degree$, i.e., the circle 90\degree\ away from the direction of motion. Positive deviations are in red and negative deviations are in blue. The size of the symbols indicates the magnitude of the velocity deviations. Circles: Mg~II, triangles: Fe~II \label{fig:resid-aitoff} }
\end{figure}

The trend exhibited in Figure~\ref{fig:VMgII} means that the cloud probed by all Components~1 is continuously being decelerated as zones progress from the anti-apex part to the apex. Alternatively,  this effect can be described as an inward motion (compression) on both sides (upstream and downstream) in the direction of propagation, and an outward motion (expansion) in the  directions perpendicular to the propagation. \\
A display of the residuals shown in Fig.~\ref{fig:VMgII} is presented in a two-dimensional representation of the sky in Figure~\ref{fig:resid-aitoff}.
This map first shows that, apart from a lack of targets in the northern part of the last quadrant (l between -90 and 0 \degree), the sight lines are more or less distributed about the sky, so that the velocity deviations are not the result of a severely lopsided distribution of the sampling directions. Most importantly, this map 
reveals that the large positive deviations (red symbols) are indeed roughly distributed along a great circle located at 90\degree\ away from the direction of motion,  which means that the expansion perpendicular to the flow vector is seen in all azimuthal angles. This is exactly what is expected if the cloud experiences  a compression in the motion direction, which naturally should produce an expansion in the perpendicular directions if the internal pressure (or volume) remains constant, similar to the deformation undergone by a free-floating balloon being squashed by a rapid push.
Whatever medium is providing pressure confinement of the cloud, e.g. the hot gas filling the Local Bubble, could also exert ram-pressure forces that create internal distortions and generally cause the material inside the cloud to decelerate.  Alternatively, perhaps the cloud is being subjected to drag forces from a magnetic field.

\subsection{A simple model for the velocity deviations\label{sec:model}}
We have expressed the mean velocity vector of the cloud in both the heliocentric and LSR frames of reference, yet, we would like to work in a fundamental reference frame, proper to the cloud's kinematics. This would enable insight into the physics of the gas, independent of the Sun or the motions of local stars. 

In this reference frame, let us envision a simple, idealized picture where the Local Cloud starts out as a spherical volume centered on the observer. Then,  we consider that  the sphere is differentially decelerated. 
 As long as the dynamical influence that decelerates the material occurs over a time interval that is comparable to or greater than the sound-crossing time, the internal pressure (and hence volume) should remain constant. 
Consequently,
the observer detects velocity perturbations from a solid-body motion that arise when the sphere tries to maintain its original volume and thus is deformed into an oblate ellipsoid that has its minor axis aligned with the deceleration vector. 

We therefore want to identify the deformation (minor) axis, which we believe defines the fundamental reference frame for the cloud's response to some external force.  

The velocity perturbations at the surface of the volume that arise from the deformation are proportional to the differences in radii of the ellipsoid and the sphere,
\begin{equation}\label{delr}
\Delta r(\theta_d)=r(\theta_d)-r_0
\end{equation}
where $r_0$ is equal to the radius of the original sphere, $\theta_d$ is the angle away from the minor axis of the ellipsoid as it deforms away from the sphere, and
\begin{equation}\label{r1}
r(\theta_d)=(a^2\sin^2\theta_d+b^2\cos^2\theta_d)^{1/2}
\end{equation}
is the radius from the center of an ellipsoid with a semimajor axis $a$ and semiminor axis $b$. Deformation velocities interior to the volume are probably smaller than those at the edge; we can assume that they scale in proportion to the distance from the observer. As we stated earlier, we adopt the restriction that the volume $V$ does not change when the sphere is deformed into an ellipsoid. This volume is given by
\begin{equation}\label{V1}
V = {4\pi\over 3}a^2b;~~ a>b~.
\end{equation}
For an eccentricity
$e=\left( 1-{b^2\over a^2}\right)^{1/2}$
we substitute the expression\begin{equation}\label{b}
b=a(1-e^2)^{1/2}~
\end{equation}
into Eq.~\ref{V1} to find that
\begin{equation}\label{V2}
V={4\pi\over 3}a^3(1-e^2)^{1/2}~,
\end{equation}
which can be transformed into an expression for $a$ that will be of use later,
\begin{equation}\label{a}
a=\left[ {3V\over 4\pi}(1-e^2)^{-1/2}\right]^{1/3}~.
\end{equation}
When we combine Eqs.~\ref{r1} and 
\ref{b}, 
we obtain
\begin{equation}\label{r2}
r(\theta_d)=a(1-e^2\cos^2\theta_d)^{1/2}~.
\end{equation}
A substitution of Eq.~\ref{a} for the term for $a$ in Eq.~\ref{r2} yields an expression for $r(\theta_d)$ that depends only on the volume $V$ and eccentricity $e$,
\begin{equation}\label{r3}
r(\theta_d)=\left({3V\over 4\pi}\right)^{1/3}(1-e^2)^{-1/6}(1-e^2\cos^2\theta_d)^{1/2}~.
\end{equation}
The volume of the original sphere before the deformation took place was given by
$V={4\pi\over 3}r_0^3~$,
which allows us to express $\Delta r(\theta_d)$ in Eq.~\ref{delr} in terms of just the volume $V$ (which is held constant) and $e$,
\begin{equation}
\Delta r(\theta_d)=\left({3V\over 4\pi}\right)^{1/3}\left[ (1-e^2)^{-1/6}(1-e^2\cos^2\theta_d)^{1/2}-1\right]
\end{equation}
Velocity deviations away from the solid-body motion reference frame should be proportional to $\Delta r(\theta_d)$ and can be expressed in function of $\theta_d$ and $e$:
 \begin{equation}\label{r4}
\Delta {\rm v}(\theta_d)=C\,\left[ (1-e^2)^{-1/6}(1-e^2\cos^2\theta_d)^{1/2}-1\right]~,
\end{equation}
where $C$ is a constant that relates changes in volume to differentials in velocity.

\subsection{Determination of the Local Cloud deformation axis \label{sec:deform}}
Following the above model, the deformation minor axis of the cloud can be found by minimizing the quadratic differences of the observed velocity deviations with respect to those expressed in equation~\ref{r4}, with the free variables being the eccentricity $e$ and the Galactic coordinates $l_d$ and $b_d$ for the origin of the angle $\theta_d$. Provided the variable $e$ is not large ($e<0.8$), the shape of the curve tracing $\Delta {\rm v}$ as a function of $\theta_d$ is invariant and its amplitude is governed only by the product of $C$ and $e$. Hence, we have in effect only one free parameter to adjust for the overall amplitude of the deviation curve. We therefore give $C$ the arbitrary fixed value of 1000 and let $e$ vary to give a best fit to the residuals. 
\begin{figure}[!htbp]
\begin{center}
\includegraphics[width=0.95\columnwidth]{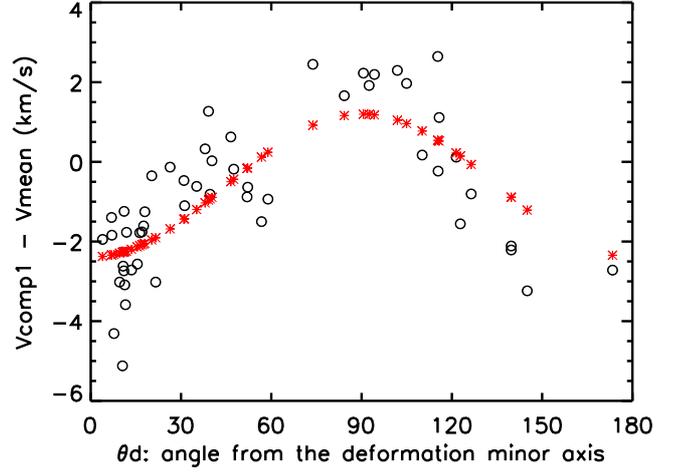} 
\end{center}
\caption{Velocity deviations relative to the mean velocity vector, plotted against the angular distance to our best fit to the cloud's minor axis  [$l_d$=174.172\degree, $b_d=-12.098$\degree].  The data points for Mg II in this plot (open circles) differ only slightly from those shown in Fig.~\protect\ref{fig:VMgII} because the deformation axis pointing toward ($l_d,~b_d$) that defines $\theta_d$ is not far from the apex of motion in the direction ($l_m,~b_m$) (see Table~\protect\ref{tab:velvectors}). The red asterisks show the model predictions.
\label{fig:resid-theta1} }
\end{figure}

With the mean velocity vector derived in Section~\ref{sec:velvectors}, marked $^{b}$ in Table~\ref{tab:velvectors} and creating the observed velocity deviations plotted in Figure~\ref{fig:VMgII}, we obtain the following direction for the deformation minor axis: \[l_d=174.17 \pm5.25 \degree; \,\,\,b_d=-12.10\pm4.21\degree\] 
and an (arbitrary) eccentricity of 0.085.
This axis defines the direction that represents a fundamental reference frame for the cloud's internal movement as it becomes distorted, so we now plot in Figure~\ref{fig:resid-theta1} the velocity deviations against the angular distance $\theta_d$ from this axis, together with the expected deviations in the model described in the previous section. 

 The match is remarkable given the simplicity of the model.
We conclude that our model of a decelerated and squashed cloud represents a reasonable first approximation to the internal kinematics of the matter surrounding the Sun. 

Note, however, that this model is most  certainly an oversimplification. 
For example, if the displacement of the gas varies in proportion to the distance from the center of the cloud, we would expect center-to-edge velocity gradients 
to create enhanced velocity dispersions (b-values) along  lines of sight roughly parallel or perpendicular to the ellipsoid's minor axis. We do not see such an effect.  We note, however, the existence of a high random dispersion in the measured b-values.  For instance, among the sight lines midway between the major and minor axes, i.e., where the b-values are expected to be  least affected by the cloud's deformation, their outcomes have a standard deviation equal to $0.7\,{\rm km~s}^{-1}$.  Given that these random errors probably arise either from the absorption profile fits or may be dominated by higher order, small scale disturbances within the cloud, it may be overly optimistic to expect that trends in the $b$-value measurements would reflect the velocity gradients that we would anticipate seeing in our highly simplified model. 

On the premise that the cloud is interacting with a motionless, external medium, or magnetic field, we may consider that the minor axis found for the spatial deformation of the cloud is very likely to indicate a fundamental direction for the cloud, independent of the chosen reference frame.  We therefore proceed with the investigations on the cloud properties by adopting the just-derived minor axis as the reference direction for the cloud. In the following discussions, we will retain the notation $\theta_d $ for the angular distance away from the minor axis of the distortion, but with the convention that $\theta_d=0$ is in the direction closest to the motion apex direction that corresponds to $\theta_m=0$. 

\section{Shape, density, and size of the Local Cloud \label{sec:density}}

We point out that the reference axis we have derived in the previous section is the minor axis that describes the velocity deviations and has no relation to the real shape of the cloud.  Even though we have used the initially-spherical cloud approximation to 
help us define our kinematical model, the cloud does not have to be perfectly regular in its shape.

The question of the shape of the cloud, or its extent in every direction around the Sun, can be addressed in an approximate manner by examining the column densities of Components~1. In effect, in a preliminary assumption that the cloud has  a homogeneous density, metal abundance, and ionization everywhere, and provided the cloud does not extend beyond the targets, the column density for a given species should be proportional to the extent of the absorbing material along that line of sight. 
\begin{figure}[!htbp]
\includegraphics[width=1.0\columnwidth]{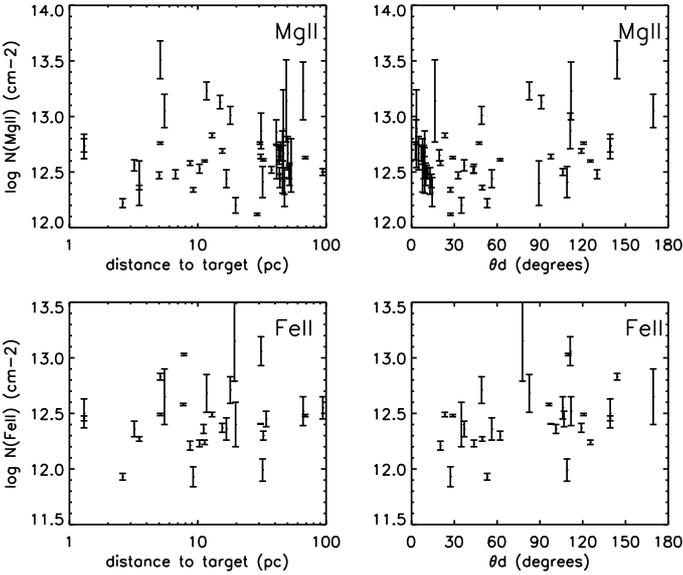}
\caption{Mg~II and Fe~II column densities against distance and against the angular distance to the deformation minor axis. 
\label{fig:col-dens-theta} }
\end{figure}

When Mg~II and Fe~II column densities are plotted as a function of distance (left-hand portion of Figure~\ref{fig:col-dens-theta}), we note a high dispersion of the data, but there is no notable increase with distance at any scale. When column densities are plotted as a function of $\theta_d$ (right-hand portion of Figure~\ref{fig:col-dens-theta}), a slight trend of $N$ increasing with $\theta_d$ is notable, however, the dispersion remains the main characteristic. We conclude  that the Local Cloud must be small because values of N do not show a general trend to increase with distance, and the large dispersion suggests that it is probably very irregular in shape. 

We noted above that the previous discussion of the shape of the cloud was valid if the cloud is homogeneous in density and abundance.
Density and extent are difficult quantities to estimate, in part because their determinations are linked. 
\cite{Wood2005} provide a good way to determine the neutral gas 
density and thus the extent of the cloud  by providing a means of identifying sight lines that are continuously filled with the cloud. This can be done 
with the information on the presence of 
an  astrosphere around the target star \citep{Linsky.Wood1996,Wood2004}. An astrosphere, analogous to the heliosphere around the Sun, is the result of the interaction of the stellar wind with neutral atoms in the surrounding interstellar matter. For a favorable geometry, the detection of an astrosphere implies that the target star is embedded within  neutral (or at least partially-neutral) gas. 
 
 \cite{Wood2005} use the results of the H~I L$\alpha$ line analyses in 59 lines of sight shorter than 80 pc and show that an astrosphere is detected for about 60\% of the sight lines shorter than 10 pc. They conclude that most of the ISM within 10 pc is filled with warm, partially neutral gas.  \cite{Wood2005} also show that $n({\rm H~I})$ varies in the local ISM and in particular in the G-cloud direction at least between 0.038 and 0.098 cm$^{-3}$. 
 
In order to derive further information on the density and cloud extent  in the context of our study, we consider  the  sample of H~I measurements  of  \cite{Wood2005} restricted to the "Components~1", i.e., the component selected in every sight line as having the velocity closest to that expected for the CHISM, as was done for the Mg~II and Fe~II sample in Section~\ref{sec:select}. 

We expect that for those targets that show an astrosphere and only one Mg~II component, it is likely  that  the Local Cloud  fills up the entire line of sight to the star.  For these cases, $N$(H~I)/$d$ should yield a good estimate of the volume density within the cloud. 
If we consider only such targets, namely $\alpha$ Cen, $\epsilon$ Eri, 36 Oph and $\xi$ Boo, 
 we are left with four relatively secure $n({\rm H~I})$ measurements of  0.098, 0.076, 0.038 and 0.040 ${\rm cm}^{-3}$, giving  an error-weighted mean of $n({\rm H~I})=0.053\,{\rm cm}^{-3}$  for the Local Cloud, in excellent agreement with the value of $n({\rm H~I})=0.055\pm0.029 \,{\rm cm}^{-3}$ derived from measurements of the composition of interstellar pick-up ions and anomalous cosmic rays in the solar system by \cite{Gloeckler2009}.
 
 If  we adopt the mean value $n({\rm H~I})$=0.053~cm$^{-3}$
and estimate the extent of the cloud in any direction by dividing $N$(H~I) by $n({\rm H~I})$, we get a mean extent of 
 8.8~pc with a rather large dispersion of 7.2~pc.

\begin{figure}[!htbp]
\includegraphics[width=1.0\columnwidth]{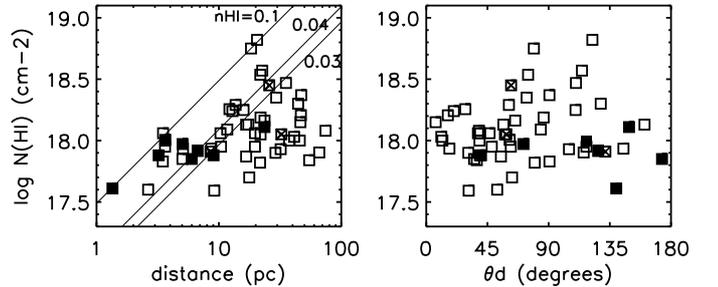}
\caption{
 H~I column densities of H~I Components~1 from the  \cite{Wood2005} sample. The filled squares indicate targets presenting evidence for an astrosphere and only one H~I component, i.e., sight lines completely filled with the Local Cloud, where N/d is supposed to give a real estimate of density. The  crossed squares indicate targets where the presence of the astrosphere has been reported but is uncertain. Diagonal lines show lines of constant densities $n({\rm H~I})=0.03,\,0.04,\,0.1\,{\rm cm}^{-3}$ . 
\label{fig:distanghi} }
\end{figure}
A few other interesting conclusions can be drawn from the examination of 
the Components~1 sample of  H~I measurements,  
shown in 
Figure~\ref{fig:distanghi}, against distance (left-hand plot) and  against  $\theta_d$ (right-hand plot):\\ 
- All data points from 0 to 9~pc lie between the two curves $n({\rm H~I})=0.03\,{\rm cm}^{-3}$ and $n({\rm H~I})=0.1\,{\rm cm}^{-3}$. Thus,  if   we consider that the first 9 parsecs are mostly filled with partially neutral gas, as suggested by  \cite{Wood2005} from the frequency of astrospheres, then its density varies from  0.03 to 0.1\,cm$^{-3}$, which extends only slightly below the bottom limit for the densities derived above from our four more secure targets.\\
 - For most targets located beyond 15 pc  the 
 column densities lie to the right of the curve of minimum observed density $n({\rm H~I})=0.03\,{\rm cm}^{-3}$,
showing that the Local Cloud does not extend up to the star. This means that in most sight lines the cloud does not extend more than  15~pc away from the Sun.\\
- A few targets situated around 20~pc have column densities compatible with the range of securely-derived density values, i.e., from 0.038 to 0.098~cm$^{-3}$, showing that the Local Cloud may extend up to 20~pc in those sight lines.  All these sight lines  are found from   $\theta_d $~=~60\degree\ to $\theta_d $~=~120\degree, i.e., roughly in the direction of the cloud deformation major axis. This elongation is consistent with what one would expect from the kinematical behavior discussed in Section~\ref{sec:model}.
\\
- For no direction indicated by $\theta_d $ do we find column densities that are particularly low, which indicates that
 there is no region where the  cloud is particularly thin, i.e., the Sun does not seem to lie near the edge of the cloud in any direction. Considering individual
measurements, we calculate that the  minimum distance of the cloud boundary from the Sun is 1.3~pc if we use the largest density value and 4.7~pc if calculated with the lowest density value.

To conclude this section, we can say that the H~I results thus favor a picture of the Local Cloud that fills up most of the space in the first 9~pc and that extends further than 15~pc  in a few directions that are approximately perpendicular to the deformation minor axis. 
 Its H~I density is always less than 0.1~cm$^{-3}$ and seems to vary in the different lines of sight by a factor 3, with a mean value of 0.053~cm$^{-3}$. There is no evidence for  a monotonic trend of n(H~I) with our $\theta_d $ angle.  In no direction is the Sun closer than 1.3 pc from the cloud boundary. 
\section{Mg~II and Fe~II abundance relative to H~I. }
\begin{figure}[!htbp]
\includegraphics[width=1.0\columnwidth]{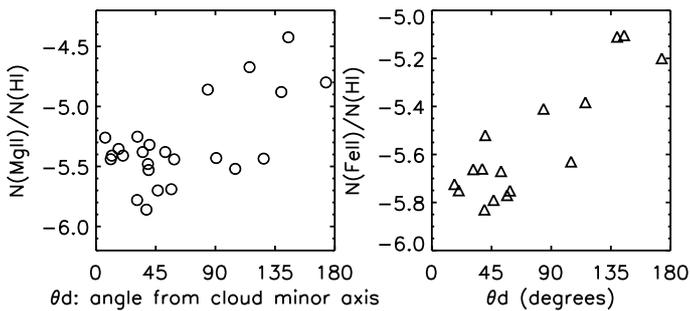}
\caption{N(Mg~II)/N(H~I) and N(Fe~II)/N(H~I) for sight lines in common to both H~I and Mg~II, Fe~II samples, plotted against $\theta_d$, the angle from the cloud's deformation minor axis pointing near the apex of motion.
\label{fig:abundance} }
\end{figure}

To derive abundances, we have restricted our sample to sight lines where measurements exist of both the H~I and the metal lines. Mg~II and Fe~II column density ratios to hydrogen are plotted against $\theta_d$ in Figure~\ref{fig:abundance}. A clear correlation is visible indicating that there is an abundance gradient within the cloud. 

We have checked that the gradient is not due to an analysis bias: Since the H~I profiles are broader than the Mg~II and Fe~II profiles, it is possible that one H~I component  encompasses several Mg~II and Fe~II components, all of which might be blended within the H~I profile. In this case, the ratio N(Mg~II)/N(H~I) would be underestimated. We have therefore repeated Figure~\ref{fig:abundance} by replacing in each case N(Mg~II) or N(Fe~II) by the sum of column density in Component~1 and up to two of the closest components to it. The trend is a bit worse, but the overall depletion levels do not change much, showing that this bias, if it exists, is not significant. 

To explain the gradient observed in the cloud
we can  invoke either ionization or depletion effects. 
\subsection{An ionization gradient in the cloud ?}
We know that the LISM is partially ionized. \cite{Jenkins2000} have shown that its ionization is maintained by a strong extreme-ultraviolet flux from nearby stars and hot gases rather than an incomplete recovery from a past, more highly ionized condition. Therefore, because of the shielding effect by the cloud material itself, the ionizing flux is rapidly decreasing with cloud depth, and an ionization gradient is expected from the surface of the cloud where it is exposed to the ionizing radiation toward the center of the cloud.

\begin{figure}[!htbp]
\includegraphics[width=1.0\columnwidth]{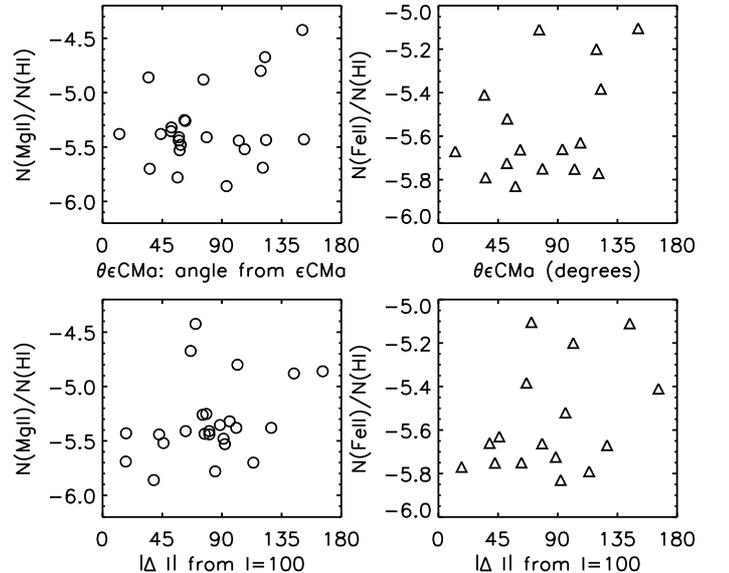}
\caption{Mg~II and Fe~II abundances against (top panels) the angle away from the dominant ionizing star $\epsilon$~CMa and (bottom panels) $|\Delta{\rm l}|$ from l=100\degree, representing a proxy for the 975\AA\ radiation field of \cite{Frisch.Andersson2012}. 
The correlations visible in Figure~\ref{fig:abundance} are not apparent here, showing that there is no evidence that the abundance variations are related to ionization effects.
\label{fig:ecma} }
\end{figure}

\cite{Vallerga1998} has shown that at the solar system position the stellar EUV ionization field above 504 \AA\  is dominated by the star $\epsilon$ CMa located at the Galactic coordinates $l=239.8$, $b=-11.3$ and to a lesser extent $\beta$ CMa at $l=226.1$, $b=-14.3$. 
\cite{Welsh2013} have recently estimated the size of the Str\"{o}mgren spheres around all white dwarfs inside the Local Bubble and also conclude that their contribution to the ionization of the LISM is secondary compared to that from the CMa stars. As a result, the hydrogen ionization in the cloud must be greatest on the side facing the direction of $\epsilon$~CMa ($\theta_d$~=~64.6\degree) and $\beta$~CMa ($\theta_d$~=~50.9\degree).
However, the top panels of Fig.~\ref{fig:ecma} demonstrate that when abundances are plotted against angular distance to the direction of $\epsilon$~CMa, there is no apparent correlation. Therefore, the abundance variation in the cloud does not seem to be related to the main ionizing source. 

\cite{Frisch.Schwadron2013} claimed that they had evidence for a correlation between the 975 \AA\  radiation field at the Sun and the Mg~II and Fe~II abundances, based on nine lines of sight in the LIC.  We looked for such a correlation  in our sample of 57 Components~1 by plotting (bottom panels of Figure~\ref{fig:ecma}) the Mg~II and Fe~II abundances against Galactic longitude, taken as a very approximate indicator for the strength of their 975 \AA\  radiation field, according to Figure 12 of \cite{Frisch.Andersson2012} that shows that the field increases smoothly from l=100\degree, where it is minimum, to l=280\degree, where it is maximum. With this definition for the field strength (based simply on $\Delta l$, not an angular distance from any point on the sky), we still find no correlation.  

\cite{Jenkins2004} showed that in low column density regions where photons with energies above 13.6~eV can penetrate the cloud, an element can be more or less ionized than H, depending on its relative $\Gamma(X)/\alpha(X)$ compared to the same for hydrogen, with $\Gamma(X)$ being the photoionization rate of the element $X$ and $\alpha(X)$ its recombination rate. For a specific example, we note that argon atoms should have a higher ionization fraction than H when a gas is partly photoionized.  \cite{Jenkins2000} have shown that it is the case in the LISM. Figure 1.1 of \cite{Jenkins2004}  shows that the same can be true for Fe~II: if the gas is irradiated by photons more energetic than the He ionization edge, $N$(Fe~II)/$N$(H~I) could underestimate the true value of Fe/H, while the converse is true for $N$(Mg~II)/$N$(H~I). 
At wavelengths below the HeI ionization edge, the ionization field is dominated by three white dwarfs Feige~24($\theta_d$~=~34.6\degree), HZ43 ($\theta_d$~=~109.0\degree), and G191-B2B ($\theta_d$~=~28.9\degree) \citep{Vallerga1998}, all of which are situated in the same quadrant as $\epsilon$~CMa. At locations that are removed from the edge of the cloud near $\epsilon$~CMa, the hydrogen opacity should block the softer photons from $\epsilon$~CMa and the field would be more dominated by the harder radiation from WD stars, which would tend to make the apparent [Fe/Mg] more severely negative than the real ratio. 
Therefore, if ionization effects were dominating the abundance variations in the cloud, we would expect to find an apparently enhanced depletion for iron and an apparently weaker depletion for magnesium as we retreat away from directions in the general direction of $\epsilon$~CMa. 
This is not what we observe, since no trend in abundances are visible on the  top panels of figure~\ref{fig:ecma}.
Furthermore, as we will see in the next section, the magnesium depletion in the Local Cloud appears to be larger than expected, instead of weaker.

We conclude that there is no evidence that supports the notion that the observed abundances are related to any gradient in the ionization field. 

\subsection{A depletion gradient in the cloud\label{sec:depletion}}
The lack of evidence for ionization effects prompts us to examine the prospect that the abundance variations could be related to a variation in the total gas-phase abundances of magnesium and iron. 
In Table~\ref{tab:densities}, we consider three zones of the cloud defined by their angular distance $\theta_d$ to the apex along the cloud's minor axis. In a rough sense these three zones represent the front (apex), the middle, and the rear (anti-apex) of the cloud. 
\\
The first three lines in Table~\ref{tab:densities}   list the mean average ion densities derived in the sightlines shorter than 15~pc.  We note no decrease  in $\bar{n}_{H~I}$ with $\theta_d$, while a noticeable, although not quite statistically significant, trend exist for $\bar{n}_{Mg~II}$ and $\bar{n}_{Fe~II}$. This attests that the apparent abundance variations are not caused by variations in the HI density.
\\
{\footnotesize
   \begin{table}[!h]
  \caption[]{Logarithms of the mean densities (in $cm^{-3}$) for Mg~II and Fe~II, their abundances relative to H, and the mean depletions, averaged in three different cloud regions  defined by the angle  $\theta_d$  from the apex direction on the cloud minor axis. 

         \label{tab:densities}}
         \begin{tabular}{lccc}
            \hline
  in log &   $\theta_d$ < 60\degree  &  60\degree<$\theta_d$ < 120\degree  &  $\theta_d$ >120\degree\\
            \hline
$\bar{n}_{Mg~II}$$^a$  &  -6.78$\pm$0.21 &  -6.50$\pm$0.16	 &  -6.26$\pm$0.50\\
$\bar{n}_{Fe~II}$$^a$   & -7.07$\pm$0.31 &   -6.80$\pm$0.34     &  -6.53$\pm$0.44\\
$\bar{n}_{H~I}$$^a$     	&-1.39$\pm$0.24 &  -1.23$\pm$0.15      &  -1.32$\pm$0.19 \\
\hline
\hline
${{\rm N}_{Mg~II}/{\rm N}_{H~I}}^{b}$                        &-5.48$\pm$0.18 &    -5.17$\pm$0.38&     -4.88$\pm$0.42\\
${{\rm N}_{Fe~II}/{\rm N}_{H~I}}^{b}$                            &-5.71$\pm$0.09&    -5.44$\pm$0.13&   -5.14$\pm$0.05\\ 
\hline
$[Mg/H]^{b,c}$                                 &-1.10$\pm$0.18&   -0.79$\pm$0.38&  -0.50$\pm$0.42\\
$[Fe/H]^{b,c}$                                 & -1.25$\pm$0.09&   -0.98$\pm$0.13&   -0.68$\pm$0.06\\
   \hline
\end{tabular}\\
$^a$  {\footnotesize \it  calculated on all sight lines shorter than 15 pc  }\\
$^{b}$  {\footnotesize \it calculated on sight lines common to both H~I and Mg~II  Fe~II samples}\\
$^{c}$ {\footnotesize \it  depletion relative to the solar abundances:  log(Mg/H)$_{\sun}$=-4.38$\pm0.02$   log(Fe/H)$_{\sun}$=-4.46$\pm0.03$ \citep{Lodders2003}
}
  \end{table}
}
To estimate the mean abundances within the three zones of the cloud we calculate the average of the column density ratios in the sight lines where both H~I and Mg~II or Fe~II are measured.
We derive depletion values, noted $[X/H]$ and listed in the last two rows of the table, by comparing the abundances to the solar abundances of \cite{Lodders2003}. 
We note an important and consistent depletion gradient inside the cloud: both elements are  about four times ($\sim$0.6~dex)  more depleted in the front of the cloud than in the rear of the cloud. 
A correlation of Mg~II and Fe~II depletion with the angle to the LIC downwind direction
had already been noted by RL08, which they interpreted mainly as a physical difference between the LIC and the G Cloud. The present correlation with the angle to the cloud minor axis shows that the depletion is gradually changing from the downwind to upwind directions inside what we consider here to be a unique cloud. 
It is noteworthy that the depletion values exhibit an almost constant difference of  $0.15-0.19$~dex for Mg relative to Fe in all three cloud regions.  

The overall depletion level is significant, especially toward the downwind direction.
Compared to the minimum ISM depletion values [X$_{gas}$/H]$_0$ defined by \cite{Jenkins2009},i.e., [Mg$_{gas}$/H]$_0= -0.270 \pm$0.030,
the magnesium depletion value derived in the front of the cloud is high, much higher than values usually found for such diffuse gas. 
We have worried about possible errors  that could result from saturation of the strong Mg~II lines. To check for this  effect, we have redone Figure~\ref{fig:abundance} after taking out the most saturated lines ($\tau$(h) >4). This resulted in the  elimination of only the 2 high Mg~II abundance points around $\theta_d =$ 80 and 120\degree, with the consequence that the magnesium abundance is unchanged for $\theta_d$<60\degree\ and $\theta_d$>120\degree.

In conclusion, it appears that the high depletion of magnesium observed in the cloud cannot be attributed to errors caused by saturation.

\cite{Jenkins2009} had examined depletion levels of the LISM in sight lines toward white dwarfs in the Local Bubble and concluded that moderately strong depletions can be found in the Local Bubble. He  also 
 obtained a significant dispersion in the depletion values, although we see no coherent trend with  $\theta_d$ in his sample, which is not surprising since the WD samples extend to material well beyond the local cloud. 

A high depletion value for magnesium had already been noticed by \cite{Kimura2003} who studied the gas and dust elemental abundances in the LIC: they found  magnesium and iron depletions around, respectively, -1.1 and -1.4, i.e., values close to our values in the  $\theta_d$<60\degree\ region, and noted that the magnesium depletion was close to that  in the general cold neutral medium (CNM) value adopted from \cite{Welty1999},  while the iron depletion was close to the general warm neutral medium (WNM) value from \cite{Sembach2000}. Due to the position historically attributed to the LIC, all their  LIC lines of sight but two have $\theta_d$ between 20\degree\ and 70\degree, and none have $\theta_d$>120\degree. This explains that they found for the LIC a value close to what we find for the $\theta_d$<60\degree\ region. 
Despite the high depletion value for magnesium, \cite{Kimura2003} show that the elemental composition of dust in the LIC implied by the different element depletions  resembles that of cometary dust in the solar system, as expected if the comets were formed from similar interstellar dust.
\cite{Kimura2003} takes the high depletion found in the LIC (or the front of the cloud for us)  as evidence that grains have not been (completely) destroyed. \cite{Frisch2005} derive the same conclusion from  the detection in the solar system of large grains of interstellar origin. 

The  gradient that we observe shows, however, that the situation is different in the rear part of the cloud, where the gas-phase abundance is significantly higher, which indicates that   Mg and Fe atoms have been returned to the gas phase after grain destruction. We note that our depletion values in the rear region of the cloud are close to the WNM depletion for magnesium, and even less depleted than the minimum value for iron.

We conclude that the rear part of the local cloud seems to be undergoing or to have undergone a grain-destructive process. 

It is tempting to relate this  to an on going interaction that would  also be responsible for the velocity deviations observed in the cloud: the cloud  squashing would be due to an acceleration from the back rather than a frontal collision.  We will note  in Section~\ref{sec:othervel}, however,  a relative lack of secondary components in the area of the cloud anti-apex, which disfavors the idea that significant interactions are taking place in that area. Therefore, this leads us to favor the interpretation that the metal abundance reflects past conditions that may be related to the origin of the cloud, as  proposed by \cite{Kimura2003}, who suggested that the LIC is  one of the cloudlets expelled from the interaction zone between the Local Bubble and the Loop~I superbubble \citep{Breitschwerdt2000}. 
\section{Nature of the other absorption components}

 Out of 59 lines of sight, 35 (59\%) show
at least one other velocity component in addition to  Component~1. 
Thirteen lines of sight (22\%) show at least two other velocity components, one line of sight shows  three, and one shows five other components.
 In the following part of this paper, we will examine the properties of the other components that can be derived from the observations, to try and pinpoint their nature. We will, in particular, examine their characteristics relative to  Components~1.
In a given sight line, we number the components  in order of increasing velocity shift $|V-V_{comp1}|$ relative to Component~1, i.e., when there are more than one other component in the line of sight, Component 2 is the component closest to Component~1 in velocity etc... 
\subsection{The Local Cloud is the dominant absorber in the LISM \label{sec:dominant}}
\begin{figure}[!htbp]
\begin{center}
\includegraphics[width=0.8\columnwidth]{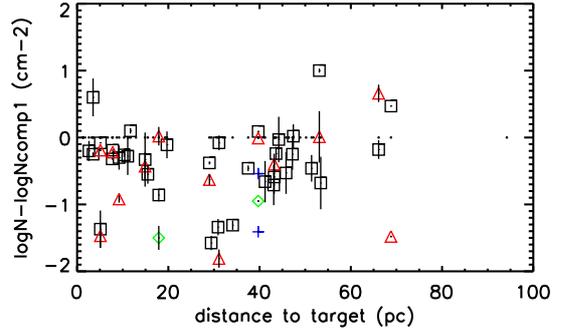}
\end{center}
\caption{Column density ratios (in log) between the other components and their corresponding Component~1: log N/Ncomp1 against distance. The column density is that of Fe~II when available and Mg~II otherwise. 
The dots are components 1. They show the position of all lines of sight. The other components, numbered in order of increasing $|\Delta V|$ relative to Component~1 are as follows: Black squares are components 2, red triangles components 3, green diamonds components 4 and blue crosses components 5 and 6. 
\label{fig:othercomp-N} }
\end{figure}

Figure~\ref{fig:othercomp-N} shows the  column density ratios between the other components and their corresponding Component~1, expressed as log N/N$_{comp1}$.  Since the determination of N(Fe~II) is less affected by saturation problems, we consider Fe~II column densities when they are available and Mg~II otherwise. %
%
As already noted in Figure~\ref{fig:velocity-domine},  Component~1 is the strongest component in terms of column density in all  lines of sight but  four. Figure~\ref{fig:othercomp-N}  shows that three of the  four sight lines where Component~1 is not dominant correspond to some of the most distant stars of the sample: $\upsilon$~Peg at d=53.1pc, $\iota$ Cap at d=66.1pc, and G191B2B at d=68 pc ($\upsilon$~Peg  had been removed from the Components~1 sample in Section~\ref{sec:residuals} because of its kinematical uncertainty but has been kept for completeness in the  sample of other components). The strongest component in  these three sight lines is therefore most probably due to  gas situated far away from us on the line of sight.\\

For the one short sightline toward 61 Cyg A (3.5 pc), where Component~1 appears as non dominant, 
 the profile fitting is very uncertain. In the words of the authors \citep{Wood.Linsky1998}:
``The low S/N of the data and the highly blended nature of the two components prevent us from deriving a unique fit to the data without first reducing the number of free parameters. We address this problem by forcing both components to have the same Doppler parameter.'' Since the optical depths of the components (according to the result of the fit) are 2.2 and 8.6, the column densities could certainly be quite different if other b-values were chosen, to the point that the column density ratio between the two components could be reversed.
Therefore,  it is not  excluded   that the Local Cloud also has the strongest column density  in this sight line.

In the  H~I sample of \cite{Wood2005}  the Component~1 is also the largest component in all sight lines shorter than 50 pc, which corroborates the above finding, even if the broad Lyman $\alpha$ profiles can easily hide unresolved components and makes the H~I count of components less reliable.

In conclusion,  what we have kinematically identified as the Local Cloud  represents the bulk of the matter in the first 50  pc. Within this distance,  the other components are truly secondary in nature.

\subsection{The gas responsible for most of the other absorbing components is very nearby and could be part of the Local Cloud
\label{sec:nearby}} 

\cite{Redfield.Linsky2004} had shown that the average number of absorbers per line of sight was relatively constant from 10 pc to 60 pc, and concluded that the LISM clouds are concentrated at no more than 15 pc from the Sun. 
Figures~\ref{fig:othercomp-N} or \ref{fig:othercomp-vdif} show that the number of sight lines with two components and the  number of sight lines with three components are  actually the same for all target distances, even for the shortest target distances: multiple components already exist for the shortest sight lines, with two components  detected toward stars located at  2.7, 3.5, and 3.6 parsecs, and three components  detected in two 5pc-long sight lines.  
It follows that  other components exist in sight lines that could be  shorter than the extent of the Local Cloud. Moreover, since longer sight lines do not show more components than very short ones, we conclude that, except for a few cases like the strong components in the long sight lines mentioned above, the gas responsible for the other components could be  as close as the Local Cloud.

As a consequence of the above conclusion, 
we suggest that  some of the secondary components could be  generated inside the Local Cloud: they could 
represent secondary velocity groupings inside the cloud. 
This possibility had already been brought forward by \cite{Gry.Jenkins2001} to reconcile the fact that  the LIC seemed to be extended all the way out to Sirius while, along this same line of sight,  a second, apparently embedded component is observed.
Supporting this picture was the finding that this same component in the line of sight toward $\epsilon$ CMa is less ionized than the LIC, whereas it should be much more ionized if it were exposed to the ionizing flux of  $\epsilon$ CMa without the shielding of the LIC.

 If our interpretation is correct, we surmise that some of these components could be perturbations belonging to the Local Cloud.
 In the next section, we will examine the collective properties of velocities and locations in the sky, much as we had done for Component~1 to identify it as the principal coherent cloud that surrounds us.
\subsection{ Categorizing the other components 
\label{sec:othervel}} 
The method usually adopted to analyze a sample of absorption  components  around the Sun is to look for groups of components that could define coherently-moving clouds from their velocities and their location on the sky (\citealt{Lallement1986,Frisch2002},~RL08). RL08 have identified not less than 15 clouds in the first 15 pc, some of which they find to be interacting. We do not adopt the same strategy since  i)  this would probably lead us to define too large a number of clouds compared to the number of available velocity components, meaning that each cloud  would be defined with too few directions (our observation sample contains fewer lines of sight and fewer absorption components than that used by RL08 because we restricted it to Mg~II and Fe~II) ; ii) the derived solution could never be unique, given the limitation of the  wavelength precision of the data and the fact that we now know from our analysis of the Local Cloud that clouds do not behave as perfectly rigid bodies ; and iii) finally for the reasons discussed earlier, 
we think that it is appropriate  to look for coherent motions relative to the Local Cloud.

Therefore, instead of trying to identify the secondary components by their heliocentric or LSR velocities, we feel that it is worthwhile to examine the velocities of these secondary clouds relative to the Local Cloud.
%
\begin{figure}[!htbp]
\begin{center}

\includegraphics[width=1.0\columnwidth]{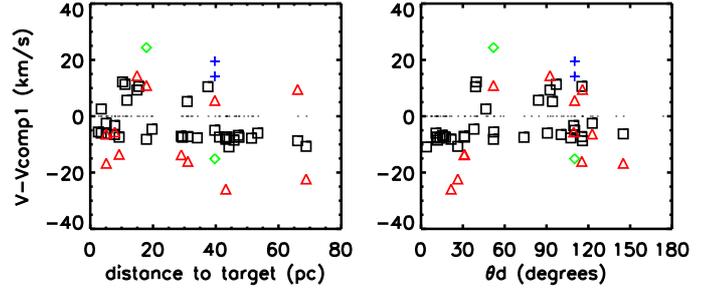} 
\end{center}
\caption{Velocity shifts of the other components relative to Components~1, plotted against  distance and the angle away from the cloud's deformation axis $\theta_d$.  
For most sight lines the velocities are Mg~II's. In those four cases where there is no Mg~II data, Fe~II velocities are used.  
The meanings of different symbols are the same as those in figure~\ref{fig:othercomp-N}. Notice the horizontally aligned symbols just below zero: these are what we refer to as the "Cetus Ripple" components.
\label{fig:othercomp-vdif} }
\end{figure}
Figure~\ref{fig:othercomp-vdif}  shows the velocity shifts ${\rm v}-{\rm v}_{comp1}$ of the other components relative to that of Component~1  and it does indeed reveal 
an interesting consistency 
in the distribution of  the other components. In particular we note that:\\
1)  The velocity shifts for all of the secondary clouds are confined within the interval $-26$ and $+24~{\rm km~s}^{-1}$ and have a mean modulus value of 9.7 ${\rm km~s}^{-1}$ with a dispersion of 10.6 ${\rm km~s}^{-1}$. 
\\
2) Only 30\% of all secondary  components have a positive shift relative to Component~1. These red-shifted secondary components are found only away from the cloud motion axis, at 45\degree<$\theta_m$<120\degree.
 They are found in  the northern hemisphere  as well as in a band in the southern hemisphere between b$\simeq -30\degree$ and b$\simeq -60\degree$\\
3) Most important, about 
half of all sight lines include a component presenting a 
velocity shift in a small range between $-12$ and $-4~{\rm km~s}^{-1}$ relative to Component~1. These components also represent half of all secondary components (26 out of 50). 
These 26 components have a mean velocity shift of 
$-7.2~{\rm km~s}^{-1}$ with an rms dispersion of 1.5 ${\rm km~s}^{-1}$ 
 relative to the Local Cloud component. 
\begin{figure}[!htbp]
\begin{center}
\includegraphics[width=0.9\columnwidth]{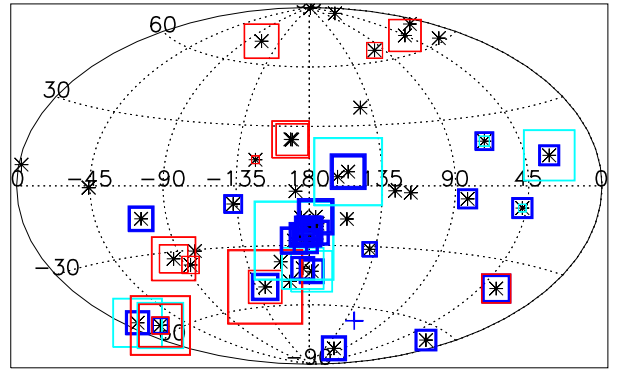} 
\end{center}
\caption{Map of the other components showing their velocity shift relative to Component~1. Black asterisks signal the position of all targets.  Blue/red squares mark components other than Components 1 with negative/positive shifts, and the size of the symbols scales with the amplitude of the shift.   Dark blue symbols indicate the  components having a uniform velocity shift of 
$-7.2\pm$1.5 ${\rm km~s}^{-1}$ relative to the Local Cloud (i.e., the Cetus Ripple); light blue symbols represent the other blue-shifted components. 
The blue cross indicates the rough center of the hemisphere occupied by the Cetus Ripple  components. 
\label{fig:aitoff}}
\end{figure}
According to Figure~\ref{fig:othercomp-vdif},  such a component is found at all values of $\theta_d$ angles, however,  the map shown in Figure~\ref{fig:aitoff}, where they are marked as 
dark blue squares of similar size,
shows that they occupy only one half of the sky 
that mainly covers the southern Galactic hemisphere. For reasons that we will justify later, we refer to them collectively as the "Cetus Ripple." It might be relevant to note that in every line of sight where blue-shifted components are detected, one of them belongs to the Cetus Ripple.
\subsection{A common origin for the components identified with the Cetus Ripple?}
\begin{figure}[!htbp]
\begin{center}
\includegraphics[width=0.6\columnwidth]{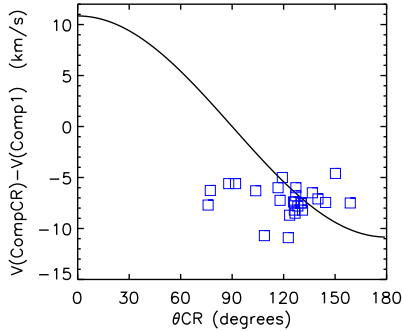}
\end{center}
\caption{Velocity shifts of Cetus Ripple (CR) components  relative to Component~1, and the cosine law corresponding to  their fitted mean vector 
plotted against the angle $\theta_{\rm CR}$ to the mean vector apex [$l_{\rm CR}=304.1\degree$, $b_{\rm CR}= +66.6\degree$].\label{fig:fit-dark-blue} }
\end{figure}

Given that  the Cetus Ripple components have been distinguished by their velocity relative to Component~1, we have  examined whether they could represent a coherent motion in the Local Cloud's reference frame, i.e., we have looked for a velocity vector whose projection in all sight lines would fit their velocity shifts with respect to Component~1. The best least-squares fit resulted in a vector with an amplitude of 10.8 $\pm$0.4 ${\rm km~s}^{-1}$ and an origin in the direction (l=124$\pm$4\degree ; b=$-67\pm3$\degree) located in the constellation of Cetus. Within the limitations of our uneven sampling of the sky, this direction is roughly consistent with the center of the hemisphere showing the targets with dark blue symbols in the plot shown in Figure~\ref{fig:aitoff} (this center point is marked with a blue cross on the map). However,  we see in Figure~\ref{fig:fit-dark-blue} that the velocities of these components relative to the Local Cloud fail to follow a  cosine curve. Despite the fact that the angles $\theta_{\rm CR}$ away from our best attempt to define an apex have been optimized to yield the best fit to a cosine relationship, the Cetus Ripple velocities do not show any real trend aside from a constant shift with respect to Components~1. We conclude that the attempt to find a coherent linear motion failed and that these  components exhibit motions that are more characteristic of an implosion.

%
%

With their uniform blue shift relative to Component~1, the Cetus Ripple components resemble  the signatures of a shock progressing toward the interior of the Local Cloud. 
This possibility had been invoked by \cite{ Jenkins1984,Jenkins2009b} as a possible consequence of the pressure imbalance that was believed to exist between the hot Local Bubble gas and the embedded diffuse clouds \citep{Bowyer1995}.
Indeed, if a  shock wave  was progressing radially with a given velocity from the outside of  the cloud toward the center, at the exact center of the cloud we would see an absorption with a constant velocity shift relative to that of the cloud. The Sun is probably not located at the center of the cloud, and the cloud contour is irregular, creating the small dispersion observed in the Cetus Ripple velocity shifts.

 If the cloud were totally immersed in the hot gas in all directions,  we would expect to see the shock in all lines of sight. We only see it in half of the sky, and this could mean that the hot gas is present only on one side of the cloud or reached the cloud from one side. This could be consistent  with a recent event at the origin of the hot gas, as in the case of a supernova blast wave approaching from one direction. 

Therefore, we consider that a common origin for the Cetus Ripple components, in the form of an inward-moving shock wave  inside the Local Cloud, is a plausible hypothesis.

\subsection{A plausible model for the imploding shock\label{sec:shock_model}}

The group of components that we identify with the Cetus Ripple exhibits an average 
velocity shift of ${\rm v}_{\rm obs} of -7.2\,{\rm km~s}^{-1}$ with an rms dispersion of $1.5\,{\rm 
km~s}^{-1}$ relative to the main Local Cloud.  The internal velocity dispersions 
exhibit $b$-values that for the most part fall within the range 2 to $4\,{\rm 
km~s}^{-1}$.  We can estimate the H~I column densities for these components by 
multiplying their values of $N$(Mg~II) and $N$(Fe~II) by the Local Cloud's average 
values of $N({\rm H~I})/N({\rm Mg~II})$ and $N({\rm H~I})/N({\rm Fe~II})$, 
and after doing so we find an average value $\log N({\rm H~I})=17.8$ with an rms 
dispersion of 0.4$\,$dex.  We now examine whether or not these properties are 
consistent with our interpretation that we could be viewing a shock converging 
upon us from the outer portions of the Local Cloud.

Within the Local Cloud, there is a magnetic field directed approximately toward the 
Galactic coordinates $l=38\degree$, $b=23\degree$ \citep{Frisch.Andersson2010}, which is 
roughly perpendicular to the center of the hemisphere that exhibits the Cetus Ripple 
components.   This orientation indicates that it is appropriate for us to examine the 
properties of a shock that, for the most part, propagates in a direction that is 
perpendicular to the field lines (i.e., the field direction is parallel to the surface of the 
shock front).   We will now explore the consequences of having either one of two 
extremes for the pre shock field strength, drawing upon some values quoted in the 
literature for the estimates of the field in the immediate environment surrounding 
the heliosphere: $B_1=2.2\mu$G \citep{Ben-Jaffel2012}  and $3.8\mu$G 
\citep{Ratkiewicz2008}. (Here, we adopt a subscript 1 to denote pre 
shock conditions and, in later derivations, subscripts 2 and 3 will respectively apply 
to the immediate post shock gas and gas further downstream from the front that 
has cooled somewhat.)  We assume that the pre shock gas has a representative 
electron density $n_1(e)=0.10\,{\rm cm}^{-3}$ \citep{Redfield.Falcon2008} and that 
virtually all of the electrons arise from the ionization of hydrogen.  With our average 
value for the volume number density of neutral hydrogen $n_1({\rm 
H~I})=0.05\,{\rm cm}^{-3}$, we expect the total particle density consisting of 
hydrogen, accompanying helium atoms, and electrons to equal $n_1=0.265\,{\rm 
cm}^{-3}$.  The gas mass density  $\rho_1=3.53\times 10^{-25}\,{\rm g~cm}^{-
3}$.  \cite{Redfield.Linsky2004b} determined an average temperature for the local 
gas equal to 6680$\,$K, and we will adopt this value for our characterization of the 
pre shock gas temperature $T_1$.  

For a shock to exist, its velocity ${\rm v}_s$ must exceed the magnetosonic speed, given by
\begin{equation}\label{v_ms}
{\rm v}_{\rm ms}=(c_{\rm s}^2+{\rm v}_{\rm A}^2)^{1/2}~,
\end{equation}
where the sound speed $ c_{\rm s}=(\gamma p_1/\rho_1)^{1/2}$ and the 
Alfv\'{e}n speed is ${\rm v}_{\rm A}=B_1/(4\pi \rho_1)^{1/2}$.  For 
$B_1=[2.2,\,3.8]\mu$G and the gas physical properties stated above, this minimum 
speed is ${\rm v}_{\rm ms}=[15.0,\,21.0]\,{\rm km~s}^{-1}$.   Gas would appear to move 
at a velocity near v$_s$ only if it had a chance to cool to some temperature 
considerably lower than the immediate post shock temperature $T_2$ and did not have its compression limited by a magnetic field (this is a 
regime that is often referred to as an ``isothermal shock,'' but is not really 
isothermal).  Instead, we view only the immediate post shock gas at an elevated 
pressure $p_2$ and temperature $T_2$, along with some gas downstream from the 
front that has undergone a moderate decrease in temperature to $T_3$.  Given 
sufficient time, this postshock gas will ultimately approach the thermal equilibrium 
value $T_{\rm eq}$ for the compressed gas.  In the paragraphs that follow, we will 
present some numerical examples of shock properties that can produce absorption 
features that match the data for the Cetus Ripple, in order to show that the shock 
interpretation is plausible.  

We can calculate the velocity v$_2$ of the immediate post shock gas in the reference 
frame of the shock front after we have computed a compression ratio 
$x_2=\rho_2/\rho_1=B_2/B_1$ across the shock front and recognize that mass flux
conservation requires that ${\rm v}_1\rho_1$ ahead of the front must equal ${\rm v}_2\rho_2$ 
behind the front.  For an Alfv\'{e}n Mach number $M_{\rm A}={\rm v}_s/{\rm v}_{\rm A}$ and 
ordinary Mach number $M={\rm v}_s/c_{\rm s}$, the shock front compression ratio is 
given by
\begin{equation}\label{x_2}
x_2={2(\gamma+1)\over D+[D^2+4(\gamma+1)(2-\gamma)/M_{\rm 
A}^2]^{1/2}}~,
\end{equation}
where
\begin{equation}\label{D}
D=\gamma-1+2/M^2+\gamma/M_{\rm A}^2
\end{equation}
\citep{Draine.McKee1993}, and the ratio of specific heats for constant pressure to 
constant volume is $\gamma=5/3$.  In the reference frame of the pre shock gas, the post shock gas starts out with a (negative) 
velocity
\begin{equation}\label{v2prime}
{\rm v}_2^\prime={\rm v}_2-{\rm v}_s={\rm v}_s[(1/x_2)-1]
\end{equation}
and gradually speeds up as the gas cools and undergoes further compression.
Therefore, to match our observed velocity of
$-7.2\,{\rm km~s}^{-1}$ for the Cetus Ripple (in the Local Cloud reference frame, 
but neglecting possible reductions resulting from projection effects), the magnitude of ${\rm v}_2^\prime$ should be 
 slightly less than $7.2\,{\rm km~s}^{-1}$.
We iterate with trial 
values of ${\rm v}_s$ and converge on acceptable values ${\rm v}_s=[19.8,\,26.0]\,{\rm 
km~s}^{-1}$ and $x_2=[1.44,\,1.32]$ for $B_1=[2.2,\,3.8]\mu$G.

From the requirement that 
momentum is conserved across the front, 
\begin{equation}\label{momentum_conserv1}
\rho_1{\rm v}_1^2+p_1+{B_1^2\over 8\pi}=\rho_2{\rm v}_2^2+p_2+{B_2^2\over 8\pi}~,
\end{equation}
we expect to find a thermal pressure behind the front given by
\begin{equation}\label{p_2}
p_2=p_1+\left(1-{1\over x_2}\right)\rho_1{\rm v}_1^2+{B_1^2(1-x_2^2)\over 8\pi}~,
\end{equation}
and thus it follows that the post shock temperature is
\begin{equation}\label{T_2}
T_2={p_2\over x_2n_1k}={T_1\over x_2}+{\rho_1{\rm v}_1^2\over n_1k}\left({1\over 
x_2}-{1\over x_2^2}\right) +{B_1^2\over 8\pi n_1 k}\left({1\over x_2}-x_2\right)~.
\end{equation}
For the two values of ${\rm v}_s$ given above, we find that
 $p_2/k=[3336,\,2875]\,{\rm cm}^{-3}$K
(compared to $p_1/k=1770\,{\rm cm}^{-3}$K), and $T_2=[8745,\,8220]\,$K.  We 
note that these post shock temperatures are not sufficient to collisionally ionize 
hydrogen \citep{Gnat.Sternberg2007}.

At any point downstream from the front, we 
again apply the requirement that momentum is conserved to find that
\begin{eqnarray}\label{momentum_conserv2}
\rho_2{\rm v}_2^2+p_2+{B_2^2\over 8\pi}&=&\rho_3{\rm v}_3^2+p_3+{B_3^2\over 8\pi} 
\nonumber\\
&=&{\rho_2{\rm v}_2^2\over x_3} + p_2{T_3\over T_2}x_3 + {B_2^2\over 8\pi}x_3^2~,
\end{eqnarray}
where $x_3=\rho_3/\rho_2$ is the additional compression that occurs when the gas 
cools.  A root to the cubic equation
\begin{equation}\label{cubic}
{B_2^2\over 8\pi}x_3^3+p_2{T_3\over T_2}x_3^2 - 
\left(\rho_2{\rm v}_2^2+p_2+{B_2^2\over 8\pi}\right)x_3 + \rho_2{\rm v}_2^2=0
\end{equation}
reveals the value of $x_3$.  For the post shock temperatures mentioned earlier, the 
rate at which the gas cools is a very strong function of temperature.  For the case  
$B_1=2.2\,\mu$G, we find that as the gas cools from $T_2=8745\,$K to $T_3=T_2-
2000\,{\rm K}=6745\,$K,  a value still well above $T_{\rm eq}\lesssim 5490\,$K, 
the average cooling rate $\langle\Lambda_{\rm H}\rangle\approx 1.3\times 10^{-
25}{\rm erg~s}^{-1}{\rm (H~nucleus)}^{-1}$ and $x_3=1.23$ at the end point.  Our 
choice for $T_3$ may seem arbitrary, but later this value for the truncation of the 
post shock flow (or the lifetime of the shock) will be seen to give a good duplication 
for the average properties of the Cetus Ripple.

The cooling is approximately isobaric, hence the cooling time is given by
\begin{equation}\label{t_cool}
t_{\rm cool}={5\over 2}k\Delta T{1.1n_2({\rm H}) +n_2(e)\over n_2({\rm H})\langle\Lambda_{\rm H}\rangle}=0.30\,{\rm Myr}~,
\end{equation}
where $n_2({\rm H})$ refers to the density of both H~I and H~II behind the front 
(strictly speaking, this should be an average of the densities in both zone 2 and zone 
3, instead of just zone 2.)  Over the time interval $t_{\rm cool}$, the shock moving at 
$19.8\,{\rm km~s}^{-1}$ traverses a distance of 6.1$\,$pc and accumulates in its 
post shock zone $\log N({\rm H~I})=17.97$, which is very close to our observed 
representative value $\log N({\rm H~I})=17.8$ for the Cetus Ripple components.  
The velocity ${\rm v}_2^\prime$ starts at $-6.0\,{\rm km~s}^{-1}$ just behind the shock
front and increases in magnitude to ${\rm v}_3^\prime=-8.6\,{\rm km~s}^{-1}$ when the temperature 
reaches $T_3=6745\,$K, giving an average velocity ${\rm v}_{\rm obs}=-7.2\,{\rm km~s}^{-1}$ that we found for the 
Cetus Ripple.  The value of ${\rm v}_3^\prime$ is calculated using Eq.~\ref{v2prime} after substituting $x_2x_3$ for $x_2$.
The expected velocity spread $\Delta {\rm v}^\prime={\rm v}_2^\prime - {\rm v}_3^\prime$ is only $2.6\,{\rm km~s}^{-1}$, so the 
absorption features should not be broadened by an unacceptably large amount.

For our other alternative, $B_1=3.8\,\mu$G, somewhat different conditions are 
required to match the properties of the Cetus Ripple components.  Here, we must 
propose that the post shock temperature drops by only 1000$\,$K below 
$T_2=8220\,$K, which results in $x_3=1.07$.  Since both the shock compression factor $x_2$ and temperature 
$T_2$ in this case are slightly lower that what we had in the previous example, we 
find that the cooling rate is diminished to a value $\langle\Lambda_{\rm 
H}\rangle\approx 1.0\times 10^{-25}{\rm erg~s}^{-1}{\rm (H~nucleus)}^{-1}$.  
For $\Delta T=1000\,$K, we find that $t_{\rm cool}=0.19\,$Myr, which means that 
the shock will travel over a distance equal to 5.0$\,$pc and processes $\log N({\rm 
H~I})=17.89$.  For this model, we find a mean value 
${\rm v}^\prime=6.9\,{\rm km~s}^{-1}$ 
and a velocity spread $\Delta {\rm v}^\prime=1.2\,{\rm km~s}^{-1}$.

In reality, the true nature of the shock may differ from our picture, either globally or from one location to another.  For representative values of the input parameters $\rho_1$ and $T_1$ we derived an outcome for $x_2$ that conformed to our measured value ${\rm v}^\prime\approx 7.2\,{\rm km~s}^{-1}$.  While average values these two input quantities in the Local Cloud have been reasonably well defined by observations, deviations probably occur over scales that may be much smaller than the overall dimension of the cloud.  For any fixed value for the strength of the shock's driving mechanism, changes in the pre shock conditions may alter ${\rm v}^\prime$ by a small amount (by about $1\,{\rm km~s}^{-1}$), but the outcome for the physical state of the post shock gas could be very different.  For instance, if $\rho_1$ is decreased by a factor of 2 and $T_1$ is increased by the same factor to maintain the same thermal pressure, we find that for $B_1=2.2\mu$G we obtain $T_2=17,500\,$K, at which point collisional ionizations will be important in increasing the ionization level of the H, Mg, and Fe \citep{Gnat.Sternberg2007}.  This might appear to be an attractive solution for the disappearance of the Cetus Ripple in certain directions, but the observations do not support the notion that $\rho_1$ is lower in this sector of the sky.

By evaluating the solutions for two different magnetic field strengths, we have demonstrated that the magnitude of the inferred pressure enhancement that drove the shock $p_2/k$ shows only minor deviations on either side of a value $3000\,{\rm cm}^{-3}$K.  While this outcome seems reasonably robust against uncertainties in field strength, we cannot guarantee that the initial field direction is parallel to the shock front, as we have assumed in our calculations.  Solutions for oblique shocks are complex and can lead to interpretations that are difficult to confirm observationally.  Likewise, we cannot be certain that the field strength near the cloud boundary is the same as the value just outside the heliosphere.

The fact that we see a constant velocity shift of the Cetus Ripple over a large range of angles in the sky (Fig.~\ref{fig:fit-dark-blue}) indicates that the shock was probably driven primarily by an enhancement of the external thermal pressure rather than from a $\rho {\rm v}^2$ momentum transfer from some specific direction.  On the premise that this external pressure elevation arises from gas at $T\sim 10^6\,$K that is currently in the Local Bubble [but see alternate possibilities expressed by \cite{Welsh.Shelton2009}], we find that after applying the calibration of the Rosat soft X-ray flux measurements to emission measures shown by \cite{Snowden1997}, a $3\sigma$ upper bound for the local diffuse X-ray emission $1.1\times 10^{-6}{\rm counts~s}^{-1}{\rm arcmin}^{-2}{\rm pc}^{-1}$ determined by \cite{Peek2011} yields a pressure $p/k=3700\,{\rm cm}^{-3}\,$K, which is only slightly above our derived pressure $p_2/k\approx 3000\,{\rm cm}^{-3}$K.

Our shock interpretation is supported by the findings of \cite{Gry.Jenkins2001} for the sight line toward $\epsilon$ CMa. 
 By comparing the two ratios Mg~II/Mg~I and C~II$^*$/C~II, they found an elevated  temperature $T> 8200\,$~K for the component that had a radial velocity equal to $-7\,{\rm km~s}^{-1}$ with respect to the strongest component with $5700 < T < 8200\,$K.  In our current interpretation, the former represents the post shock gas column, while the latter belongs the pre shock material. At the lower bound for the temperature of the post shock gas, however, they found that $n(e) = 0.09\,{\rm cm}^{-3}$, which is about equal to $n(e) = 0.08$ that they found at the upper bound for temperature in the pre shock gas.  Superficially, we might say that there is a lack of evidence for any compression in the post shock gas, but it is possible that the average for $n(e)$ over the entire sight line in the strong component is higher than that of the gas in the immediate vicinity of the shock front.
 
 We acknowledge that our proposal regarding the nature of the Cetus Ripple components is somewhat  speculative.  
Nevertheless, as suggested by the kinematical properties shared by these components, we still retain the picture that some single physical phenomenon may be responsible for creating them.  A single shock wave is probably the simplest phenomenon that is consistent with this picture. 

 \subsection{Nature of the remaining components that are neither Components 1 nor Cetus Ripple Components \label{sec:otherother}}
 The remaining components constitute a minor fraction of the matter.
 We have seen in Section~\ref{sec:dominant} that Components~1 are the dominant components in all sight lines shorter than 50 pc. In fact, Components~1 account on average for 70 \% of the total MgII and FeII column density for sight lines shorter than 50 pc. Adding the contribution of the Cetus Ripple components, then we find that the Components 1 and the Cetus Ripple components together account for  85 \% of the total column density within 50 pc of the Sun.

Figure~\ref{fig:othercomp-vdif}  indicates that the remaining  components happen to be blueshifted relative to the Local Cloud for  sight lines close to the  direction of motion, at $\theta_d < $30\degree 
 and $\theta_d > $110\degree, and redshifted for sight lines comprised between $\theta_d$=30 and 115\degree away from the direction of motion. This is not sufficient to pinpoint their nature. Within the framework of our proposal for the nature of the Cetus Ripple components, these extra components could arise from secondary shocks or small velocity waves induced by turbulent motions.
Redshifted components could alternatively be formed by gas  escaping from the cloud. This is made more plausible 
by the observation that the presence of redshifted components is usually associated with the absence of a Cetus Ripple component.  This could suggest that one side of the cloud might not be as strongly constrained by the surrounding hot gas.

As already mentioned in Section~\ref{sec:dominant} some of the extra components, in particular for sight lines $> 50$ pc,  may be external clouds. We do not have enough long sight lines, however, to relate several features together and define any coherent cloud motions for them.

\section{Summary and conclusions}
We have re-examined the sample of LISM sight lines (d$\leq$100 pc) observed  at high spectral resolution in the UV lines of Fe~II and Mg~II.
\begin{enumerate}
      \item The kinematical study shows that all lines of sight contain a component, which we call Component~1, whose velocity is consistent to within $\pm 6\,{\rm km~s}^{-1}$ with the projection of the velocity vector of the material entering the heliosphere. We make the hypothesis that all manifestations of Component~1 have their origin in a unique cloud surrounding the Sun, which we call the Local Cloud. We derive the mean velocity vector that best fits the velocity of all Components~1, and check that it is in close agreement with the velocity of the interstellar flow in the heliosphere. We then find that 
      velocity residuals from this mean vector follow a special trend: they are  negatively shifted for directions close to the apex and anti-apex directions and positively shifted for directions perpendicular to the motion direction. After accounting for this trend, the remaining deviations are not higher than the velocity measurement errors. We conclude that the kinematic information from the Fe~II and Mg~II UV lines sample is consistent with the existence of a single cloud surrounding the Sun provided the cloud is not constrained to behave like a rigid body.
    \item To interpret the velocity residuals, we develop a simple model  where the Local Cloud starts out as an idealized  spherical volume centered on the observer and then is deformed into an oblate ellipsoid as it is decelerated. 
     We show that the observed velocity perturbations away from the motion of a rigid body fit well with this interpretation. From this behavior we derive the minor axis of cloud's deformation, which is  aligned with the deceleration vector. This deformation minor axis is taken as the natural reference axis for the cloud and does not depend on any frame of reference. 
     \item Following \cite{Wood2005} we derive the H~I density in the Local Cloud from the H~I measurements in a few sight lines where the cloud is known to extend to the target, owing to the presence of an astrosphere around the target. The H~I density appears to vary between 0.03 and 0.1 cm$^{-3}$ with a mean value around  0.053 cm$^{-3}$, in agreement with the measurements that apply to the immediate vicinity of the solar system. When considering all N(H~I) measurements  we find that the cloud extends  to 8.8 pc on average with a dispersion of 7.2 pc, with a few directions where it could extend up to about 20 pc and a lower limit  of 1.3 pc for the minimum cloud extent in any direction.
       \item The abundances of Mg~II and Fe~II  show a gradient with the angular distance $\theta_d$ away from the deformation axis of the cloud. This gradient does not seem to be  related to the major  ionization processes. Therefore, we conclude that Mg~II  is significantly depleted onto grains in the Local Cloud, and the depletion level decreases from the head to the rear of the cloud. 
             \item    The Local Cloud is the dominant absorber in almost all  of the sight lines, even if a second velocity component is present in 60\% of the Mg~II and Fe~II sight lines,  and a third velocity component is present  in 22\% of the sight lines.  
Since the presence and the number of secondary components are independent of the distances to the targets, and since they are present already in very short sight lines, we believe that they could be velocity groupings inside the cloud. 
Since half of the secondary components,  covering half of the sky in the southern hemisphere, are blueshifted from the Local Cloud component with a relatively constant shift of $-7.2\, {\rm km~s}^{-1}$, we suggest that they could be kinematically related to the Local Cloud.  We have named this unique collection of components the "Cetus Ripple." We can reproduce their characteristics by the presence of an implosive shock progressing radially inward toward the Sun.
We show that with  pre shock conditions corresponding  to the density, temperature, and magnetic field observed in the Local Cloud, it is possible to find a shock velocity that yields post shock conditions matching our observations for the Cetus Ripple components in terms of velocity, column density, and velocity dispersion.

              \end{enumerate}
            
 In conclusion, 
 we offer an innovative view of the LISM made of one main monolithic cloud that fills the space around the Sun out to a distance of about 9 pc. 

Small velocity displacements are interpreted as the signature of a large scale distortion undergone by the cloud perhaps in response to a differential deceleration. The elemental abundances also evolve from the cloud anti-apex to its apex, indicating that dust has been processed differentially inside the cloud along the axis of motion. Finally, we advocate that one half of the directions present evidence for a shock wave progressing toward the cloud interior, possibly created by a sudden increase of external thermal pressure.

 We stress that this picture is fundamentally different from previous models (e.g.,  \citealt{Lallement1986,Frisch2002},~RL08) where the LISM is constituted of a collection of small clouds or cloudlets that are presented as separate entities moving as rigid bodies at different velocities in slightly different directions.  In particular, in our picture, the LIC, the G cloud,  and other distinct clouds of the RL08 model, are unified in a single local cloud.
 
Although both pictures originate in the same kinematical data, they are based on very  different  fundamental assumptions: on the one hand, multiple, rigid, and homogeneous clouds and on the other hand, a single, heterogeneous cloud subject to  distortions. Differences in the initial assumptions yield substantially divergent morphologies. 
\begin{acknowledgements}
Based on observations made with the NASA/ESA Hubble Space Telescope, obtained from the data archive at the Space Telescope Institute. STScI is operated by the association of Universities for Research in Astronomy, Inc. under the NASA contract NAS 5-26555. This research has made use of the SIMBAD database, operated at CDS, Strasbourg, France.    \\   

CG is very grateful to Princeton University Department of Astrophysical Sciences, and especially to Ed Jenkins and Bruce Draine for their fantastic hospitality during the two-month visit when this work was initiated.   We appreciate suggestions for improving the paper from Bruce Draine and an anonymous referee.

\end{acknowledgements}
\bibliographystyle{aa}
\bibliography{biblio}

\begin{thebibliography}{53}
\expandafter\ifx\csname natexlab\endcsname\relax\def\natexlab#1{#1}\fi

\bibitem[{{Ben-Jaffel} \& {Ratkiewicz}(2012)}]{Ben-Jaffel2012}
{Ben-Jaffel}, L. \& {Ratkiewicz}, R. 2012, \aap, 546, A78

\bibitem[{{Bowyer} {et~al.}(1995){Bowyer}, {Lieu}, {Sidher}, {Lampton}, \&
  {Knude}}]{Bowyer1995}
{Bowyer}, S., {Lieu}, R., {Sidher}, S.~D., {Lampton}, M., \& {Knude}, J. 1995,
  \nat, 375, 212

\bibitem[{{Breitschwerdt} {et~al.}(2000){Breitschwerdt}, {Freyberg}, \&
  {Egger}}]{Breitschwerdt2000}
{Breitschwerdt}, D., {Freyberg}, M.~J., \& {Egger}, R. 2000, \aap, 361, 303

\bibitem[{{Crawford} {et~al.}(1998){Crawford}, {Lallement}, \&
  {Welsh}}]{Crawford1998}
{Crawford}, I.~A., {Lallement}, R., \& {Welsh}, B.~Y. 1998, \mnras, 300, 1181

\bibitem[{{Crutcher}(1982)}]{Crutcher1982}
{Crutcher}, R.~M. 1982, \apj, 254, 82

\bibitem[{{Dehnen} \& {Binney}(1998)}]{Dehnen.Binney1998}
{Dehnen}, W. \& {Binney}, J.~J. 1998, \mnras, 298, 387

\bibitem[{{Draine} \& {McKee}(1993)}]{Draine.McKee1993}
{Draine}, B.~T. \& {McKee}, C.~F. 1993, \araa, 31, 373

\bibitem[{{Frisch} {et~al.}(2005){Frisch}, {Gr{\"u}n}, \& {Hoppe}}]{Frisch2005}
{Frisch}, P., {Gr{\"u}n}, E., \& {Hoppe}, P. 2005, ISSI Scientific Reports
  Series, 3, 183

\bibitem[{{Frisch} {et~al.}(2010){Frisch}, {Andersson}, {Berdyugin}, {Funsten},
  {Magalhaes}, {McComas}, {Piirola}, {Schwadron}, {Slavin}, \&
  {Wiktorowicz}}]{Frisch.Andersson2010}
{Frisch}, P.~C., {Andersson}, B.-G., {Berdyugin}, A., {et~al.} 2010, \apj, 724,
  1473

\bibitem[{{Frisch} {et~al.}(2012){Frisch}, {Andersson}, {Berdyugin}, {Piirola},
  {DeMajistre}, {Funsten}, {Magalhaes}, {Seriacopi}, {McComas}, {Schwadron},
  {Slavin}, \& {Wiktorowicz}}]{Frisch.Andersson2012}
{Frisch}, P.~C., {Andersson}, B.-G., {Berdyugin}, A., {et~al.} 2012, \apj, 760,
  106

\bibitem[{{Frisch} {et~al.}(2013){Frisch}, {Bzowski}, {Livadiotis}, {McComas},
  {Moebius}, {Mueller}, {Pryor}, {Schwadron}, {Sok{\'o}{\l}}, {Vallerga}, \&
  {Ajello}}]{Frisch.Science2013}
{Frisch}, P.~C., {Bzowski}, M., {Livadiotis}, G., {et~al.} 2013, Science, 341,
  1080

\bibitem[{{Frisch} {et~al.}(2002){Frisch}, {Grodnicki}, \&
  {Welty}}]{Frisch2002}
{Frisch}, P.~C., {Grodnicki}, L., \& {Welty}, D.~E. 2002, \apj, 574, 834

\bibitem[{{Frisch} \& {M\"{u}ller}(2011)}]{Frisch.Mueller2011}
{Frisch}, P.~C. \& {M\"{u}ller}, H. 2011, Space Sci Rev, 176, 21

\bibitem[{{Frisch} {et~al.}(2011){Frisch}, {Redfield}, \&
  {Slavin}}]{Frisch.Redfield.Slavin2011}
{Frisch}, P.~C., {Redfield}, S., \& {Slavin}, J.~D. 2011, \araa, 49, 237

\bibitem[{{Frisch} \& {Schwadron}(2013)}]{Frisch.Schwadron2013}
{Frisch}, P.~C. \& {Schwadron}, N.~A. 2013, ArXiv e-prints, 1310.2922,
  proceedings of the 12th Annual International Astrophysics Conference, in
  press

\bibitem[{{Frisch} \& {Slavin}(2006)}]{Frisch.Slavin2006}
{Frisch}, P.~C. \& {Slavin}, J.~D. 2006, Astrophysics and Space Sciences
  Transactions, 2, 53

\bibitem[{{Gloeckler} {et~al.}(2009){Gloeckler}, {Fisk}, {Geiss}, {Hill},
  {Hamilton}, {Decker}, \& {Krimigis}}]{Gloeckler2009}
{Gloeckler}, G., {Fisk}, L.~A., {Geiss}, J., {et~al.} 2009, \ssr, 143, 163

\bibitem[{{Gnat} \& {Sternberg}(2007)}]{Gnat.Sternberg2007}
{Gnat}, O. \& {Sternberg}, A. 2007, \apjs, 168, 213

\bibitem[{{Gry} \& {Jenkins}(2001)}]{Gry.Jenkins2001}
{Gry}, C. \& {Jenkins}, E.~B. 2001, \aap, 367, 617

\bibitem[{{Jenkins}(1984)}]{Jenkins1984}
{Jenkins}, E.~B. 1984, in NASA Conference Publication, Vol. 2345, NASA
  Conference Publication, ed. Y.~{Kondo}, F.~C. {Bruhweiler}, \& B.~D.
  {Savage}, 155--168

\bibitem[{{Jenkins}(2004)}]{Jenkins2004}
{Jenkins}, E.~B. 2004, in Carnegie Observatories Astrophysics Serie, Vol.~4,
  Origin and Evolution of the Elements, ed. A.~{McWilliam} \& M.~{Rauch}, 336

\bibitem[{{Jenkins}(2009{\natexlab{a}})}]{Jenkins2009}
{Jenkins}, E.~B. 2009{\natexlab{a}}, \apj, 700, 1299

\bibitem[{{Jenkins}(2009{\natexlab{b}})}]{Jenkins2009b}
{Jenkins}, E.~B. 2009{\natexlab{b}}, \ssr, 143, 205

\bibitem[{{Jenkins} {et~al.}(2000){Jenkins}, {Oegerle}, {Gry}, {Vallerga},
  {Sembach}, {Shelton}, {Ferlet}, {Vidal-Madjar}, {York}, {Linsky}, {Roth},
  {Dupree}, \& {Edelstein}}]{Jenkins2000}
{Jenkins}, E.~B., {Oegerle}, W.~R., {Gry}, C., {et~al.} 2000, \apjl, 538, L81

\bibitem[{{Kalas} {et~al.}(2004){Kalas}, {Liu}, \& {Matthews}}]{Kalas2004}
{Kalas}, P., {Liu}, M.~C., \& {Matthews}, B.~C. 2004, Science, 303, 1990

\bibitem[{{Kamp} {et~al.}(2006){Kamp}, {Durand}, \& {Micol}}]{Kamp2006}
{Kamp}, I., {Durand}, D., \& {Micol}, A. 2006, GHRS Instrument Science Report,
  92

\bibitem[{{Kimura} {et~al.}(2003){Kimura}, {Mann}, \&
  {Jessberger}}]{Kimura2003}
{Kimura}, H., {Mann}, I., \& {Jessberger}, E.~K. 2003, \apj, 582, 846

\bibitem[{{Lallement} \& {Bertaux}(2014)}]{Lallement.Bertaux2014}
{Lallement}, R. \& {Bertaux}, J.-L. 2014, ArXiv e-prints, 1402.1977

\bibitem[{{Lallement} \& {Bertin}(1992)}]{Lallement1992}
{Lallement}, R. \& {Bertin}, P. 1992, \aap, 266, 479

\bibitem[{{Lallement} {et~al.}(1986){Lallement}, {Vidal-Madjar}, \&
  {Ferlet}}]{Lallement1986}
{Lallement}, R., {Vidal-Madjar}, A., \& {Ferlet}, R. 1986, \aap, 168, 225

\bibitem[{{Linsky} \& {Wood}(1996)}]{Linsky.Wood1996}
{Linsky}, J.~L. \& {Wood}, B.~E. 1996, \apj, 463, 254

\bibitem[{{Lodders}(2003)}]{Lodders2003}
{Lodders}, K. 2003, \apj, 591, 1220

\bibitem[{{Malamut} {et~al.}(2014){Malamut}, {Redfield}, {Linsky}, {Wood}, \&
  {Ayres}}]{Malamut2014}
{Malamut}, C., {Redfield}, S., {Linsky}, J.~L., {Wood}, B.~E., \& {Ayres},
  T.~R. 2014, \apj, 787, 75

\bibitem[{{McComas} {et~al.}(2012){McComas}, {Alexashov}, {Bzowski}, {Fahr},
  {Heerikhuisen}, {Izmodenov}, {Lee}, {M{\"o}bius}, {Pogorelov}, {Schwadron},
  \& {Zank}}]{McComas2012}
{McComas}, D.~J., {Alexashov}, D., {Bzowski}, M., {et~al.} 2012, Science, 336,
  1291

\bibitem[{{M{\"o}bius} {et~al.}(2004){M{\"o}bius}, {Bzowski}, {Chalov}, {Fahr},
  {Gloeckler}, {Izmodenov}, {Kallenbach}, {Lallement}, {McMullin}, {Noda},
  {Oka}, {Pauluhn}, {Raymond}, {Ruci{\'n}ski}, {Skoug}, {Terasawa}, {Thompson},
  {Vallerga}, {von Steiger}, \& {Witte}}]{Moebius2004}
{M{\"o}bius}, E., {Bzowski}, M., {Chalov}, S., {et~al.} 2004, \aap, 426, 897

\bibitem[{{Peek} {et~al.}(2011){Peek}, {Heiles}, {Peek}, {Meyer}, \&
  {Lauroesch}}]{Peek2011}
{Peek}, J.~E.~G., {Heiles}, C., {Peek}, K.~M.~G., {Meyer}, D.~M., \&
  {Lauroesch}, J.~T. 2011, \apj, 735, 129

\bibitem[{{Ratkiewicz} \& {Grygorczuk}(2008)}]{Ratkiewicz2008}
{Ratkiewicz}, R. \& {Grygorczuk}, J. 2008, \grl, 35, 23105

\bibitem[{{Redfield} \& {Falcon}(2008)}]{Redfield.Falcon2008}
{Redfield}, S. \& {Falcon}, R.~E. 2008, \apj, 683, 207

\bibitem[{{Redfield} \& {Linsky}(2001)}]{Redfield.Linsky2001}
{Redfield}, S. \& {Linsky}, J.~L. 2001, \apj, 551, 413

\bibitem[{{Redfield} \& {Linsky}(2002)}]{Redfield.Linsky2002}
{Redfield}, S. \& {Linsky}, J.~L. 2002, \apjs, 139, 439

\bibitem[{{Redfield} \& {Linsky}(2004{\natexlab{a}})}]{Redfield.Linsky2004}
{Redfield}, S. \& {Linsky}, J.~L. 2004{\natexlab{a}}, \apj, 602, 776

\bibitem[{{Redfield} \& {Linsky}(2004{\natexlab{b}})}]{Redfield.Linsky2004b}
{Redfield}, S. \& {Linsky}, J.~L. 2004{\natexlab{b}}, \apj, 613, 1004

\bibitem[{{Redfield} \& {Linsky}(2008)}]{Redfield.Linsky2008}
{Redfield}, S. \& {Linsky}, J.~L. 2008, \apj, 673, 283, (RL08)

\bibitem[{{Sembach} {et~al.}(2000){Sembach}, {Howk}, {Ryans}, \&
  {Keenan}}]{Sembach2000}
{Sembach}, K.~R., {Howk}, J.~C., {Ryans}, R.~S.~I., \& {Keenan}, F.~P. 2000,
  \apj, 528, 310

\bibitem[{{Snowden} {et~al.}(1997){Snowden}, {Egger}, {Freyberg}, {McCammon},
  {Plucinsky}, {Sanders}, {Schmitt}, {Truemper}, \& {Voges}}]{Snowden1997}
{Snowden}, S.~L., {Egger}, R., {Freyberg}, M.~J., {et~al.} 1997, \apj, 485, 125

\bibitem[{{Vallerga}(1998)}]{Vallerga1998}
{Vallerga}, J. 1998, \apj, 497, 921

\bibitem[{{Welsh} \& {Shelton}(2009)}]{Welsh.Shelton2009}
{Welsh}, B.~Y. \& {Shelton}, R.~L. 2009, \apss, 323, 1

\bibitem[{{Welsh} {et~al.}(2013){Welsh}, {Wheatley}, {Dickinson}, \&
  {Barstow}}]{Welsh2013}
{Welsh}, B.~Y., {Wheatley}, J., {Dickinson}, N.~J., \& {Barstow}, M.~A. 2013,
  \pasp, 125, 644

\bibitem[{{Welty} {et~al.}(1999){Welty}, {Hobbs}, {Lauroesch}, {Morton},
  {Spitzer}, \& {York}}]{Welty1999}
{Welty}, D.~E., {Hobbs}, L.~M., {Lauroesch}, J.~T., {et~al.} 1999, \apjs, 124,
  465

\bibitem[{{Witte}(2004)}]{Witte2004}
{Witte}, M. 2004, \aap, 426, 835

\bibitem[{{Wood}(2004)}]{Wood2004}
{Wood}, B.~E. 2004, Living Reviews in Solar Physics, 1, 2

\bibitem[{{Wood} \& {Linsky}(1998)}]{Wood.Linsky1998}
{Wood}, B.~E. \& {Linsky}, J.~L. 1998, \apj, 492, 788

\bibitem[{{Wood} {et~al.}(2005){Wood}, {Redfield}, {Linsky}, {M{\"u}ller}, \&
  {Zank}}]{Wood2005}
{Wood}, B.~E., {Redfield}, S., {Linsky}, J.~L., {M{\"u}ller}, H.-R., \& {Zank},
  G.~P. 2005, \apjs, 159, 118

\end{thebibliography}

{\it Note added in proof:} After the completion of our study, new observations in additional
directions were reported by \cite{Malamut2014}.  The addition of these new results should help to
refine the conclusions reported here.

\begin{appendix}
\label{appendice}
\section{Comparison with the clouds of the RL08 model.}  
\begin{figure}[!htbp]

\includegraphics[width=1.0\columnwidth]{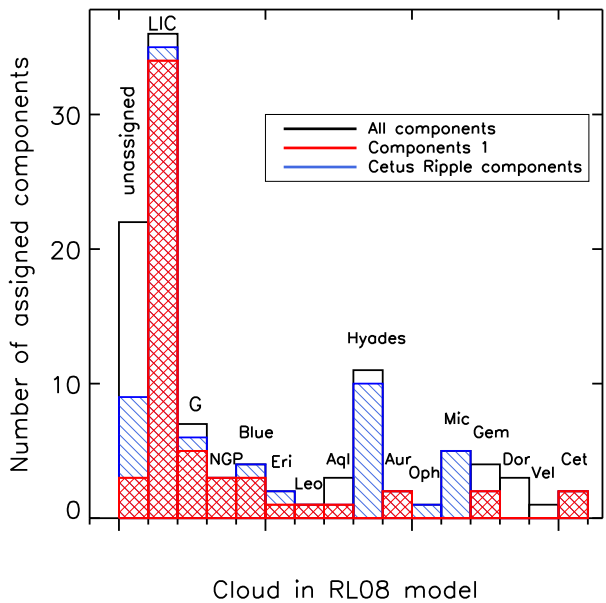} 
\caption{Histogram representing the number of components identified with a given cloud in the \cite{Redfield.Linsky2008} model (RL08). The name of the RL08 clouds are indicated on top of the bars. Black lines represent all components used in our study,  red crisscrossed areas represent the  components that we have recognized as  Components~1 and blue hatched areas represent the  components that we have called "Cetus Ripple" components. The box marked "unassigned" contains  the components listed in the "Unassigned sight line properties" Table 17 in RL08.
\label{fig:histo} }
\end{figure}
 For readers who wish to identify our Components 1 with a previous study, we show in Figure~\ref{fig:histo} 
how the absorption components of our sample were distributed among the different clouds in the RL08 model.
Note that the RL08 clouds have been defined from a larger sample since it included Ca~II ground-based data, resulting in  2.5 times more components. This explains why some of the RL08 clouds  are represented by only one or two components in the present sample. 

In red we show the distribution of our  Components~1, i.e., components we now attribute to the Local Cloud. We see, as mentioned in Section~\ref{sec:unique}, that our Components~1 had been assigned to the LIC for a majority (60\%) of them, to the G cloud for 9\% of them, to the NGP and the blue clouds, both for 5\% of them. Another 5\% of our  Components~1 had not been assigned to any cloud, and the others were distributed over six other clouds.

Out of  the 50 components other than Components~1, about half of them (26) have a velocity shift that lies within the interval $-4$ and $-12\,{\rm km~s}^{-1}$ relative to Component~1, these are the components we called the ``Cetus Ripple,''  which are consistent with the signature of a shock progressing inside the cloud toward its interior. Nine of them were assigned to the Hyades cloud, five to the Mic cloud, and a remaining five were distributed over five other clouds. 

Another way of looking at this is to note that seven of the RL08 clouds (LIC, G, NGP, Blue, Leo, Aur, Cet) include a majority of Components~1, i.e., components that are here attributed to the Local Cloud. On the other hand,  among the five RL08 clouds including no Components~1, three of them (Hyades, Mic and Oph) include almost exclusively the Cetus Ripple components, meaning that these clouds are made of matter moving toward the Local Cloud --or its interior-- at moderate speed.

A total of 22 components out of 107 in our sample  had not been assigned to any cloud in the RL08 model. In our study,  26 components are neither a Component 1 nor a Cetus Ripple component.

\end{appendix}

\end{document}